\documentclass[aps, prd, twocolumn, lengthcheck, superscriptaddress, showpacs, letterpaper, nofootinbib]{revtex4-1}
\usepackage{amsfonts}
\usepackage{amsmath}
\usepackage{amssymb}
\usepackage{xcolor}
\usepackage{graphicx}
\usepackage{subfigure}
\usepackage{longtable}
\usepackage{slashed}
\usepackage{hyperref}
\usepackage{amsthm}
\usepackage{mathrsfs}
\usepackage{color}
\usepackage{epsfig}
\usepackage{epstopdf}

\def\be{\begin{eqnarray}}\def\ee{\end{eqnarray}}
\def\bi{\bibitem}
\def\la{\langle}\def\ra{\rangle}
\def\be{\begin{eqnarray}}\def\ee{\end{eqnarray}}
\def\lsim{\mathrel{\rlap{\lower3pt\hbox{\hskip1pt$\sim$}}
     \raise1pt\hbox{$<$}}} 
\def\gsim{\mathrel{\rlap{\lower3pt\hbox{\hskip1pt$\sim$}}
     \raise1pt\hbox{$>$}}} 
\newcommand\sect[1]{\emph{#1.}---}

\allowdisplaybreaks

\begin{document}
\title{Topology change, emergent symmetries and compact star matter}

\author{Yong-Liang Ma}
\email{Corresponding author: ylma@ucas.ac.cn}
\affiliation{School of Fundamental Physics and Mathematical Sciences,
Hangzhou Institute for Advanced Study, UCAS, Hangzhou, 310024, China}
\affiliation{International Center for Theoretical Physics Asia-Pacific (ICTP-AP) (Beijing/Hangzhou), UCAS, Beijing 100190, China}

\author{Mannque Rho}
\email{Corresponding author: mannque.rho@ipht.fr}
\affiliation{Universit\'e Paris-Saclay, CNRS, CEA, Institut de Physique Th\'eorique, 91191, Gif-sur-Yvette, France }

%

\begin{abstract}
Topology effects have being extensively studied and confirmed in strongly correlated condensed matter physics. In the limit of large number of colors, baryons can be regarded as topological objects---skyrmions---and the baryonic matter can be regarded as a skyrmion matter. We review in this paper the generalized effective field theory for dense compact-star matter constructed with the robust inputs obtained from the skyrmion approach to dense nuclear matter, relying on possible ``emergent" scale and local flavor symmetries at high density. All  nuclear matter properties from the saturation density $n_0$ up to several times $n_0$  can be fairly well described. A uniquely novel---and unorthdox---feature of this theory is the precocious appearance of the pseudo-conformal sound velocity $v^2_{s}/c^2 \approx 1/3$, with the non-vanishing trace of the energy momentum tensor of the system. The topology change encoded in the density scaling of low energy constants is interpreted as the quark-hadron continuity in the sense of Cheshire Cat Principle (CCP) at density $\gsim 2n_0$ in  accessing massive compact stars. We confront the approach with the data from GW170817 and GW190425.
\end{abstract}


\maketitle

\section{Introduction}

The structure of dense nuclear matter relevant to compact stars has been investigated for several decades but still remains largely uncharted. Unlike at  high temperature, so far, the physics at high density can be accessed by neither terrestrial experiments nor lattice simulation. Recently, the observation of massive neutron stars with mass $\gsim 2.0 M_\odot$ and detection of gravitational waves from neutron star mergers provide indirect information of nuclear matter at low temperature and high density, say, up to $\sim 10$ times the normal nuclear matter density $n_0\simeq 0.16$ fm$^{-3}$~\cite{Demorest:2010bx,Antoniadis:2013pzd,Cromartie:2019kug,TheLIGOScientific:2017qsa,Abbott:2018exr,Abbott:2020uma}. These new developments offer the powerful means  to explore the nuclear matter in the interior of the compact stars, for example,  the patterns of the symmetries involved therein, what is in the core  of the stars, say, baryons and/or quarks  and a combination thereof. For recent discussions on these aspects, we suggest, e.g., Refs.~\cite{Holt:2014hma,Baym:2017whm,McLerran:2018hbz,MR-review,Strangness,Kojo:2020krb} and some relevant references therein.

The study of nuclear matter in the literature has largely relied on either phenomenological approaches anchored on density functionals or effective field theoretical models implemented with assumed QCD symmetries and degrees of freedom appropriate for the cutoff to which the theory is applicable. For finite nuclei as well as the infinite nuclear matter up to $\sim n_0$, the physics can be described very well by using the nuclear effective theory with or without pion, in addition to the nucleon~\cite{Hammer:2019poc,Holt:2014hma} (denoted as $s\chi$EFT). However, in the dense system relevant to the compact stars at $\sim 10 n_0$, the $s\chi$EFT is believed to break down. Then, to construct an effective theory for compact star matter, one should consider the following facts which may not be independent: What the interior of the star could consists of,  baryons and/or quarks, and a combination thereof? Whether and how the relevant degrees of freedom of QCD---the gluons and quarks---intervene? Are phase transitions involved in the core of the massive stars?

In the past several years, we have devoted ourselves with our collaborators to construct a general but conceptually novel nuclear effective field theory (dubbed G$n$EFT) applicable not only to the finite nuclei but also to the highly dense system relevant to the massive stars where $s\chi$EFT is considered to break down~\cite{MR-review}. The merit of the approach that we rely on is that we will have a single unified effective Lagrangian formulated in a way that encompasses from low density to high density, involving only manifestly ``macroscopic" degrees of freedom, but capturing the continuity to ``microscopic" quarks-gluon degrees of freedom---in the sense of Cheshire Cat Principle~\cite{Rho:1983bh,Goldstone:1983tu}.

When going to the density relevant to compact stars, since the nucleons are very close to each other, the effects from the hadron resonances must enter. Therefore, in addition to the nucleons and pions in $s\chi$EFT, the G$n$EFT includes the lowest-lying vector mesons $\rho$ and $\omega$ and the scalar meson $f_0(500)$. Put in terms of the degrees of freedom, the G$n$EFT can be written as
$$
\mbox{G$n$EFT}=s\chi\mbox{EFT} + \rho \mbox{~and~} \omega + f_0(500).
$$
In the model construction, in addition to the chiral symmetry, the lowest-lying vector mesons are introduced through the hidden local flavor symmetry~\cite{yamawaki,Bando:1987br,VM} and the scalar meson $f_0$ is regarded as the pseudo-Nambu-Goldstone boson of the hidden scale symmetry~\cite{crewther}. Both symmetries are not explicit in the matter free space but it seems reasonable to think that they get (partially) restored in the dense system.  At least there is nothing glaring at odds with the presently available observations.

Prior to QCD, Skyrme suggested that baryons can be described by the topology solution of a mesonic theory, skyrmion~\cite{Skyrme:1961vq}. After the arrival of QCD, it was argued that when the number of color $N_c$ is infinitely large, baryons in the constituent quark model share the same $N_c$ scaling properties as skyrmions~\cite{Witten:1979kh,manohar-largeN}. Since then, the Skyrme(-type) model~\footnote{Hereafter, for convenience, we use Skyrme model with the pion field only to represent the Skyrme model and its extensions.} has become one of the models in the study of nucleon, nuclei as well as nuclear matter~\cite{Zahed:1986qz,BReditor,MR_SkyrRev}.

In the skyrmion approach to dense nuclear matter obtained by putting skyrmions on crystal lattice, a robust observation independent of the model and crystal structure---at least what has been checked so far---is the topology change where the skyrmions with the integer winding number transit to half-skyrmions with the half-integer winding number. The density at which this takes place is denoted as $n_{1/2}$. This model-independent topology change gives rise to several interesting density dependences of hadron properties that have not been found in other approaches.

Although the Skyrme model approach can describe the nucleon, nuclei as well as nuclear matter in a unified way, it is a daunting task to put this approach into practice since the calculation depends on the efficiency of the computer and the results are valid in the large $N_c$ limit. Therefore, in practice, one resorts to chiral effective models that incorporate baryons as explicit degrees of freedom. In our G$n$EFT, we incorporate the robust characteristics of topology in the low energy constants of the model. The effect of the change of the degrees of freedom is formulated in terms of the possible topology change at a density $n_{1/2}$ encoded in the behavior of the parameters of
the G$n$EFT Lagrangian as one moves from below to above the changeover density $n_{1/2}$. After making the $V_{lowk}$ RG approach implementing the strategy of Wilsonian renormalization group flow~\cite{Vlowk}, we construct the pseudo-conformal model (PCM) of dense nuclear matter~\cite{PKLMR,PCM}(see Ref.~\cite{MR-review} for a review).

The PCM that satisfy all the constraints from astrophysics turns out to have  a peculiar feature that has not been found in any other approaches:  The sound velocity approaches the conformal limit $v_{s}^2/c^2\approx 1/3$ at the density relevant to compact stars although the trace of the energy-momentum tensor does not vanish. This is in stark contrast to the standard scenario favored in the field~\cite{Tews:2018kmu}. This conceptually novel approach predicts that the core of massive compact stars is populated by {\it confined} quasi-fermions of fractional baryon charge~\cite{Ma:2020hno},  not ``deconfined quarks" expected in perturbative QCD~\cite{evidence}.  We suggest that this phenomenon, together with the ``quenched $g_A$ problem" in nuclei, shows that hidden symmetries hidden in medium-free vacuum of QCD emerge in nuclear dynamics~\cite{gA}.

\section{Topology change and hadron--quark continuity}

It has long been discussed that in the large number of color $N_c$ limit, baryons can be regarded as topological objects---solitons, namely skyrmions. In the skyrmion approach, the dense nuclear matter can be accessed by putting the skyrmions onto the crystal lattice~\cite{Klebanov:1985qi,ParkVento,MR_SkyrRev}.
%
Here we exploit the Skyrme model with the Lagrangian connected to QCD in
the sense of Weinberg ``folk theorem'' on effective field theories
~\cite{Weinberg:1996kw}. For a development quite different in spirit from ours, we refer to, e.g., review~\cite{Adam:2015ele} and the references therein.

\subsection{Topology change}

Topology change is a novel phenomenon that has not been observed in any approach other than the skyrmion crystal approach to dense nuclear matter.

\begin{figure}[htp]\centering
\includegraphics[scale=0.25]{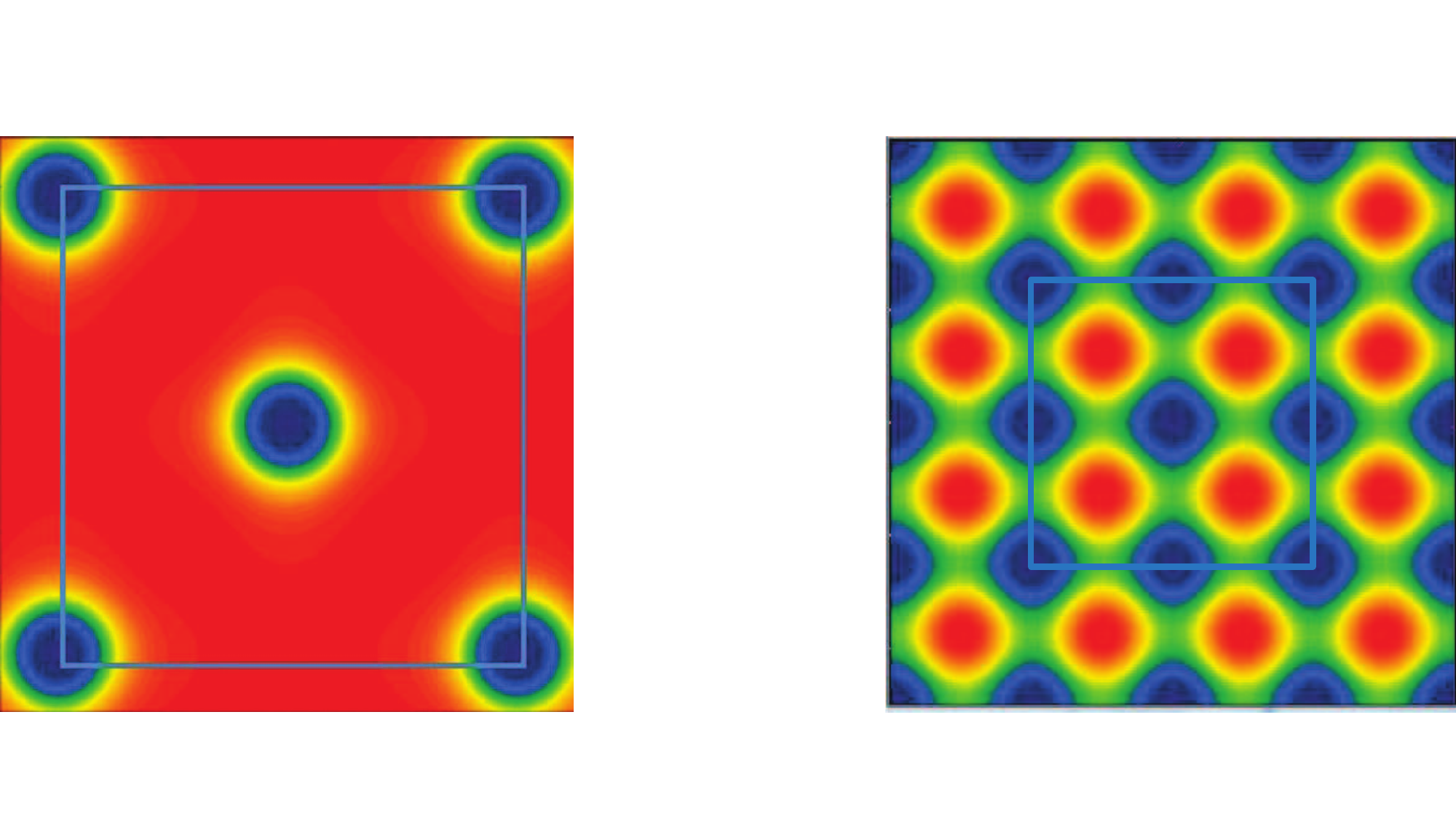}
\caption{Distribution of the baryon number density in the skyrmion (left panel) matter and half-skyrmion matter (right panel). From \cite{ParkVento}.}
\label{fig:SkyrPhaseTran}
\end{figure}

To have an intuitive idea, let us look at the distribution of the baryon number density in a specific lattice, say, face-centered cubic crystal. The distribution of the baryon number density looked along an axis is illustrated in the left panel of Fig.~\ref{fig:SkyrPhaseTran}. The winding number is $1$ if one integrates out the blue volume. Now, squeeze the system. One finds that, after a critical density $n_{1/2}$ (or equally, the crystal size $L_{1/2}$), the distribution of the baryon number density changes to the right panel of Fig.~\ref{fig:SkyrPhaseTran}.
%
%
What happens is that when the increasing matter density surpasses  $n_{1/2}$ (or the crystal size drops below $L_{1/2}$), the constituents of the matter given in the blue square  transit from winding number--$1$ objects (left panel) to winding number--$1/2$ objects, half-skyrmions (right panel). (How this happens in the numerical simulation can be seen in \cite{ParkVento} .)

In the half-skyrmion configuration, as a consequence of symmetry, the space
average of
\be
\langle \phi_0 \rangle = \frac{1}{V}\int_V d^3x \frac{1}{2}{\rm tr}U_0 \to 0,
\ee
where $U_0$ is the static configuration of the chiral field $U=\exp(2i\pi^aT^a/f_\pi)$ with $T^a = \sigma^a/2$. This means that the quark condensate $\langle\bar{q}q\rangle $ vanishes when the space is averaged. Therefore, one can use this quantity as a signal of the skyrmion--half-skyrmion transition.

It should be noted that the location of $n_{1/2}$ cannot be pined down theoretically because it is model-dependent. Since nuclear dynamics at low density can be well described by $s\chi$EFT, we set $n_{1/2}\gsim 2.0n_0$. Later, we will see that astrophysical observations observations indicate $2.0n_0\lsim n_{1/2}\lsim 4.0n_0$.

\subsection{Implications of topology change}

\sect{Chiral symmetry breaking}
In the skyrmion crystal approach to dense nuclear matter, the pion decay constant can be calculated through the axial-vector currelator~\cite{Ma:2013ela}
\be
i G_{\mu\nu}^{ab}(p) & = & i \int d^4x e^{ip\cdot x} \langle 0| T J_{5\mu}^a(x)J_{5\nu}^b(0)|0\rangle.
\ee
At the leading order of fluctuations, we can express the medium modified pion decay constant as
\be
f_\pi^{\ast2} & = & f_\pi^2 \left[1 - \frac{2}{3}\left(1 - \langle\phi_0^2\rangle\right)\right].
\ee

In the skyrmion phase, since $\langle\phi_0^2\rangle$ decreases with density, $f_\pi^{\ast2}$ decreases with density. After passing $n_{1/2}$ from below, since $\langle\phi_0\rangle = 0$ in the chiral limit, $\langle\phi_0^2\rangle \simeq 0$. Thus
\be
\frac{f_\pi^{\ast2}}{f_\pi^2} & \simeq &  \frac{1}{3}
\label{eq:fpiSkyr}
\ee
a nonzero constant although $\langle\phi_0\rangle = 0$. This argument is supported by explicit numerical calculation.

In terms of current algebra, the generalized  Gell-Mann--Oakes--Renner relation tells us~\cite{Lacour:2010ci}
\be
m_\pi^{\ast2}f_\pi^{\ast2}=m_q \langle\phi_0\rangle + \sum_{n\geq 2} F_n(\langle\phi_0^n\rangle),
\ee
where for convenience, we have kept the current quark mass. $F_n$ stands for the contribution from multiquark condensation. Since the pion mass  scales little with density, when going to the half-skyrmion matter,
\be
\frac{f_\pi^{\ast2}}{f_\pi^2} = \sum_{n\geq 2} F_n(\langle\phi_0^n\rangle) \simeq F_2(\langle\phi_0^2\rangle)  \neq 0,
\ee
which is in qualitative agreement with the result from skyrmion crystal calculation.

Equation~\eqref{eq:fpiSkyr} means that the chiral symmetry is only partially restored in the half-skyrmion matter and we are still in the Nambu-Goldstone phase. This means that the skyrmion--half-skyrmion transition is not a Landau-Ginzburg-type phase transition. Although it is not a paradigmatic phase change, in what follows, we will use the term ``half-skyrmion phase" for simplicity.

\sect{Chiral doublet structure}
It is found that when the system goes to the half-skyrmion medium, the nucleon mass becomes a density-independent constant~\cite{Ma:2013ooa}. Therefore, one can decompose the nucleon mass as
\be
m_N & = & \Delta(\langle\bar{q}q\rangle) + m_0
\label{eq:decompmass}
\ee
where $\Delta(\langle\bar{q}q\rangle)$ is the sector of the nucleon mass coming from the quark condensate which becomes zero in the half-skyrmion medium. $m_0$ is the sector of the nucleon mass independent of $\langle\bar{q}q\rangle$ and has a magnitude about $(50-70)\%$ of the nucleon mass in vacuum.
The existence of $m_0 \neq 0$ implies that there is a part of the nucleon mass that is chiral invariant.

It should be noted that, the decomposition~\eqref{eq:decompmass} can also be inferred from other approaches. The lattice calculation found that, when the chiral symmetry is unbroken, baryons are still massive and one should not expect a drop of the mass in dense medium~\cite{Glozman:2012fj}.  The same behavior was found in Ref.~\cite{Paeng:2011hy} in a renormalization group (RG) analysis of hidden local symmetric Lagrangian with baryons. Moreover, in Ref.~\cite{Motohiro:2015taa}, by using a chiral effective model with parity doubler, it was found  that, to reproduce the nuclear matter around saturation density, the nucleon mass should has a sizable chiral invariant component. So far, it is not clear to us whether $m_0$ reflects a fundamental feature of QCD or an emergent symmetry via correlations in medium as in condensed matter as indicated in this crystal calculation.

\sect{Symmetry energy}
The symmetry energy of nuclear matter $E_{sym}(n)$ which plays the most important role in the equation of state (EoS) for compact stars is not under control at the density relevant to compact stars~\cite{Chen:2015gba,Li:2019xxz}. It is given by the term proportional to $\alpha^2$ in the energy per nucleon $E(n, \alpha)$
\be
E(n, \alpha) & = & E(n, \alpha=0) + E_{sym}(n)\alpha^2 + O(\alpha^4),
\ee
where $\alpha = (N-P)/(N+P)$ with $P (N)$ being the number of protons (neutrons).

Since the symmetry energy arises from the proton-neutron asymmetry, to calculate it from the skyrmion crystal approach, the crystal lattice should be rotated through a single set of collective coordinates~\cite{Lee:2010sw}. A tedious but straightforward calculation yields
\be
E_{sym} = 1/\lambda_I,
\ee
where  $\lambda_I$ is the isospin moment of inertia.

\begin{figure}[h]\centering 
\includegraphics[scale=0.45,angle=0]{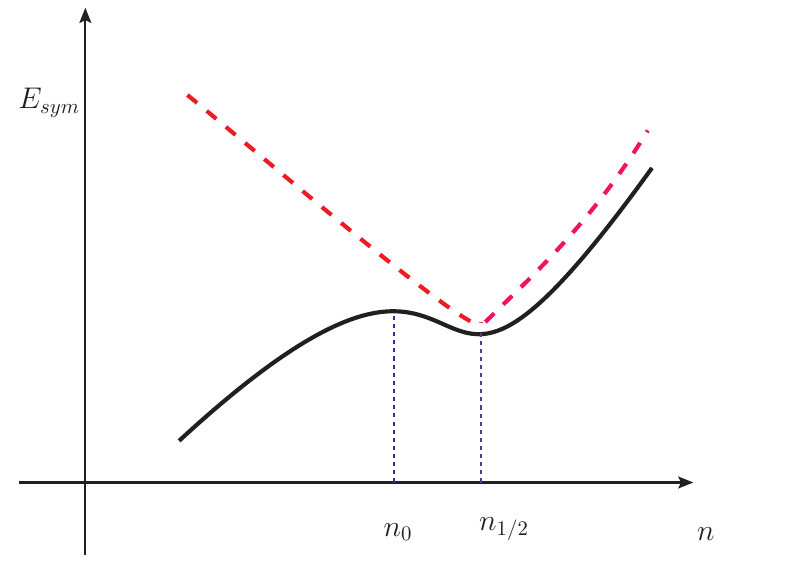}
\caption{Schematic illustration of the symmetry energy calculated by the skyrmion crystal (dashed line) and nucleon correlation corrections (solid line). }
\label{LPR-fig}
\end{figure}

The density dependence of the symmetry energy obtained from the skyrmion crystal approach is schematically plotted in dotted curve in Fig.~\ref{LPR-fig}. What is interesting is the appearance of the cusp structure locked at $n_{1/2}$, i.e., the symmetry energy first decreases with density and then increases when the density passes $n_{1/2}$. To understand the density dependence of symmetry energy, we consider the expression of $\lambda_I$~\cite{Liu:2018wgv}
\be
\lambda_I & = & \frac{f_\pi^2}{6}\left\langle 4 - \phi_0^2\right\rangle + \cdots
\label{eq:lambdaILO}
\ee
where $\cdots$ stands for the contribution from the Skyrme term and $\langle \cdots \rangle$  indicates the space average of the quantity inside. As discussed above, with the increasing of the density, $\langle \phi_0^2\rangle$ decreases to zero. So, $1/\lambda_I$, or equivalently $E_{sym}$, decreases going toward $n_{1/2}$. After $n_{1/2}$, the tendency of $E_{sym}$ is highly involved. Since at $n \gsim n_{1/2}$, $\langle \phi_0^2\rangle \approx 0$, the density dependence from the quartic term in the Skyrme model which represents massive excitations --  such as the vector mesons in the HLS models -- intervene. It gives the cusp structure.

It should be stressed that the crystal description of baryonic matter at low density cannot be reliable, so the density dependence of symmetry energy obtained at density $n \lesssim n_0$ cannot be taken seriously. The cusp structure at $n_{1/2}$ is present in nuclear correlations as is shown below in terms of nuclear tensor forces.  What is important in the skyrmion crystal calculation is that the symmetry energy decreases toward the cusp density after which it increases. We will see later that this cusp sheds  light on the medium modified-hadron properties.

\sect{Nuclear tensor force}
We have shown that, the robust characteristic in the skyrmion crystal approach is the existence of the cusp structure in the symmetry energy. A natural question is what is the implication of this cusp in G$n$EFT including nucleon as an explicit degree of freedom or equivalently, how to reproduce this cusp in G$n$EFT. To address this question, we consider the tensor force between nucleons that is mediated by one boson exchange.

The symmetry energy is dominated by the nuclear tensor force $V^T$ and can be written in the closure approximation as~\cite{Brown:1994pq}
\be
E_{sym} & \simeq & c\frac{\langle (V^T)^2\rangle}{\Delta E}.
\label{closure-sum}
\ee
Therefore, the behavior of the symmetry energy is controlled by the absolute value of the tensor force between nucleons carried by the exchanged mesons.

For the present purpose, it suffices to consider the one-pion and one-$\rho$  contributions to two-body tensor forces. The scalar meson, here dilaton, does not contribute directly at the tree level but affects indirectly on the scaling relations of the masses and coupling constants in the Lagrangian. In the non-relativistic limit, the tensor forces are given by
\be
V^T_M \left( r \right) &=& S_M \frac{f_{NM}^{\ast\,2}}{4\pi} \tau_1\, \tau_2\, S_{12}{{\cal I}(m^\ast_{M} r)} \, ,\label{tensorM}\\
{{\cal I}(m^\ast_M r)}&\equiv& m_M^\ast \left[ \frac{1}{(m_M^\ast r)^3} + \frac{1}{(m_M^\ast r)^2} + \frac{1}{3m_M^\ast r} \right] e^{-m_M^\ast r} \,,
\label{radial}
\ee
where $M=\pi,\,\rho$, $S_{\rho(\pi)} = +1(-1)$ and
\begin{equation}
S_{12} = 3 \frac{\left(\vec{\sigma}_1 \cdot \vec{r} \,\right)\left(\vec{\sigma}_2 \cdot \vec{r} \,\right) }{r^2} - \vec{\sigma}_1 \cdot \vec{\sigma}_2
\end{equation}
with the Pauli matrices $\tau^i$ and $\sigma^i$ for the isospin and spin of the nucleons with $i = 1,2,3$. The density dependence enters through the scaling parameters in the in-medium quantities marked with asterisk~\cite{Brown:1991kk}. The strength $f_{NM}^\ast$ scales as
\begin{equation}
R^\ast_M\equiv \frac{f_{NM}^\ast}{f_{NM}} \approx \frac{g_{M NN}^\ast}{g_{M NN}} \frac{m_N}{m_N^\ast} \frac{m_M^\ast}{m_M}\, ,
\end{equation}
where $g_{MNN}$ are the effective meson-nucleon couplings. What is significant in Eq.~(\ref{tensorM}) is that given the same radial dependence, the two forces (through the pion and $\rho$ meson exchanges) come with an opposite sign and therefore cancels each other.

As discussed in Ref.~\cite{PKLR}, if the hadron scales from low to high densities with no topology change, the tensor force will decrease monotonically with density. There will then be no cusp in the symmetry energy. This feature will be in conflict with what happens in Nature.

\begin{figure}[ht!]
\begin{center}
\includegraphics[width=6cm]{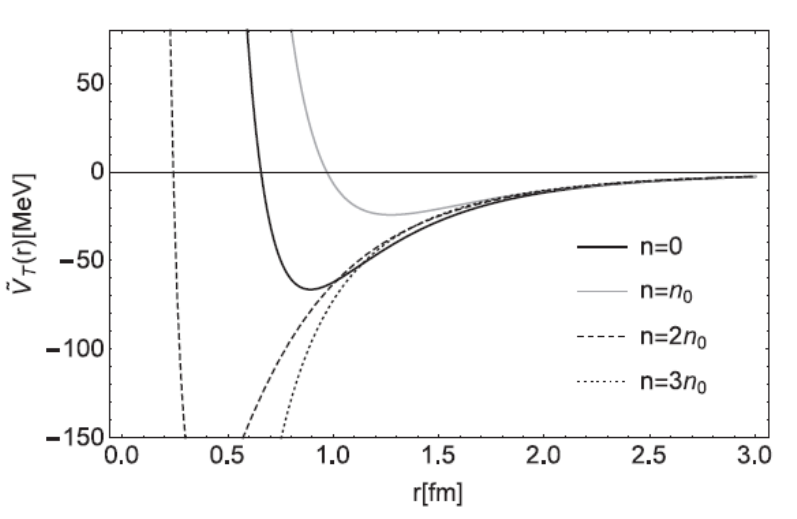}
\caption{Net tensor force $\tilde{V}^T\equiv (\tau_1 \cdot \tau_2 S_{12})^{-1} (V_\pi^T +V_\rho^T)$ with $\Phi \approx 1-0.15 n/n_0$  and $R\approx 1$ for $n<n_{1/2}$ and $R^\ast_\rho\approx \Phi^2$ for $n>n_{1/2}$ by assuming $ n_{1/2} \approx 2n_0$~\cite{PKLR}. }\label{vtwo}
\end{center}
\end{figure}

Now let us see what happens if there is the topology change at $n_{1/2}$. For illustration we take $R^\ast_\rho\approx \Phi^2$ at $n> n_{1/2}$ but with all others the same as in the case without topology change. The results are plotted in Fig.~\ref{vtwo}. It shows that the topology change effect is dramatic. Due to the cancellation between these two tensor forces, in the range of nuclear forces relevant for the
nuclear interaction, $r \gsim 1$~fm, the magnitude of the net force first decreases and then, after passing $n_{1/2}$, increases and the force from the rho meson is nearly totally suppressed. Then, from Eq.~(\ref{closure-sum}), one concludes that going toward to $n_{1/2}$ from below the symmetry energy is to drop and more or less abruptly turn over at $n_{1/2}$ and then increase beyond $n_{1/2}$. This reproduces precisely the cusp predicted in the crystal calculation. As a result, the cusp structure in $E_{sym}$---a consequence of topology change with the onset of the half-skyrmion phase---is signaling the different density scaling property of the gauge coupling from $n\leq n_{1/2}$ to $n>n_{1/2}$.

In summary, the topology change found in the skyrmion crystal approach to density nuclear matter indicates that the hadron properties, such as nucleon mass, meson masses, pion decay constant and hidden gauge coupling and so on, have different density scaling in the skyrmion and half-skyrmion phases. We will see later that, this observation has a drastic effect on the dense nuclear matter for compact stars.

We should mention here that higher correlation corrections brought in the $V_{lowk}$ renormalization flow calculation ``smoothen" the cusp in the form represented in solid line in Fig.~\ref{LPR-fig}.
\subsection{Quark-hadron continuity}

We have argued that the topology change is a robust feature in the skyrmion crystal approach to dense nuclear matter. The question is whether or how the topology change represents the ``quark deconfinement'' process in QCD. There is no clear answer at present, so we
can only offer a conjecture on how one can establish the connection in the sense of Cheshire Cat Principle (CCP) based on the chiral bag model of nucleon.

For the number of flavors $N_f \geq 2$, baryons can be described by chiral bags~\cite{Rho:1983bh,Goldstone:1983tu}. Inside the bag, the degrees of freedom are quarks and gluons, and the baryon number is carried by the quarks. Outside of the bag, mesons are the relevant degrees of freedom, and the baryon number is carried by topology in the winding number. When the bag is shrunk, all the quarks drop into the inifinite hotel and turn into skyrmions with only the Cheshire Cat smile remaining. That physics should not depend on the bag size is the CCP.

In the case of single flavor $N_f=1$, the situation is quite different because there is no $N_f=1$ skyrmion. It turns out that the baryon should be a soliton resembling a pancake~\cite{Komargodski:2018odf} or pita~\cite{Karasik:2020pwu} having a fractional quantum Hall (FQH) topology structure. There is a Cheshire Cat description for this in terms of an anomaly flow~\cite{Ma:2019xtx}.  But what is puzzling is that there are two Cheshire Cats, one involving 3D ball and the other 2D sheets. It seems very plausible that at low density baryonic matter is in skyrmions in 3D with the metastable 2D FQH pancakes/pitas suppressed. However it seems indispensable at high density that the FQH topology structure be taken into account. This is because at high density where chiral transition takes place, the vector mesons in hidden local symmetry become the Chern-Simons fields (via Seiberg-type duality). This part of the high density story is not yet understood, so we can only say that we really do not understand what happens at high density. In what we have done, we are simply assuming that the Chern-Simons fields do not figure importantly in the range of compact-star densities.  We will simply ignore this ``dichotomy problem." This aspect of the problem is discussed in \cite{D,trade-in}.

The topology change in the skyrmion crystal approach appears at the density at which the profiles of solitons overlap and the valence quarks inside the baryons rearrange to form different clusters, here configurations with baryon number--$1/2$. This picture resembles the quarkyonic matter proposed in \cite{McLerran:2007qj,McLerran:2018hbz} and the hard-core realization of the deconfinement from nuclear to quark matter phrased in Ref.~\cite{Fukushima:2020cmk}.

As mentioned above and will be discussed later, owing to the topology change implemented in the parameters in the Lagrangian of G$n$EFT, the symmetry energy $E_{sym}$, as it approaches $n_{1/2}$ from slightly below, softens and after passing  $n_{1/2}$, hardens. This generates a spike in the density dependence of the sound velocity. In Ref.~\cite{Pisarski:2021aoz}, this spike was attributed to the enhancement and then suppression of the $\omega_0$ condensate in the low and high density region. We suppose that this behavior of $\omega_0$ condensate can be naturally explained using the scale-chiral effective theory beyond the leading order scale symmetry in which not only the $\omega$ meson mass but also the $\omega$-$N$-$N$ coupling scales with density~\cite{MR-hwz}.

\section{Emergent symmetries}

After the discussion on the topology change which serves as one of the key ingredient of the PCM, let us now turn to two other essential ingredients, the hidden local gauge symmetry and hidden scale symmetry which are invisible in the vacuum of QCD. Our approach is to exploit the possible emergence of these symmetries as density increases to the regime relevant to compact stars, say, $\lsim 10 n_0$. We use these symmetries to include the higher-energy degrees of freedom---the lowest-lying vector mesons $V=(\rho,\omega)$ and the scalar meson $f_0(500)$. Here, we focus on the points directly relevant to the PCM construction,   leaving the details to~\cite{MR-review,Rho:2021zwm}

\subsection{Emergent hidden local symmetry}

To bring in the lowest-lying vector mesons $\rho$ and $\omega$ into the chiral effective theory, we adopt the strategy of hidden local symmetry (HLS)~\cite{Bando:1987br,yamawaki,VM} which at low density is gauge equivalent to nonlinear sigma model, the basis of $s\chi$EFT.

By decomposing the chiral field $U(x)$ as $U(x) = \xi_L^\dagger \xi_R$, one can introduce a redundant local symmetry $h(x)$ under which $\xi_{L, R}$ transforms as
\be
\xi_{L,R} & \to & \xi_{L,R}h^\dagger(x)
\ee
but keeps the chiral properties of $U(x)$ intact. When a chiral effective theory is expressed in terms of $\xi_{L,R}$, the gauge fields of local symmetry $h(x)$---$V(x)$---enter the theory. After higssing the gauge symmetry, the gauge fields $V(x)$ obtain masses. In HLS, the field content depends on the symmetry $h(x)$. If one chooses $h(x) \in SU(2)\times U(1)$, one can identify $V=(\rho, \omega)$ with $\rho \in SU(2)$ and $\omega \in U(1)$. It is assumed that the kinetic terms of $V_\mu(x)$ can be generated by underlying dynamics of QCD or quantum corrections, thus $V_\mu(x)$ become dynamical gauge bosons~\cite{yamawaki}. Compared to other approaches of vector mesons, with HLS, one can establish a systematic power counting  by treating the vector mesons on the same footing as the Nambu-Goldstone boson, pions~\cite{VM}.

Now, come back to the nuclear matter.  At low density where the nucleons are far from each other, the vector mesons are massive objects and can be exchanged between them. Using the equations of motion of the vector mesons, their effects are accounted for as a two-pion exchange effect, i.e.,  one-loop contribution in s$\chi$EFT. The question is in whst sense the vector mesons can be regarded as hidden local gauge fields.  The Suzuki's theorem~\cite{suzuki} states that ``when a gauge-invariant local field theory is written in terms of matter fields alone, a composite gauge boson or bosons must be inevitably formed dynamically."  If we assume the ``vector manifestation (VM)"~\cite{VM,Harada:2000kb} that $m_\rho^2 \propto f_\pi^2 g_\rho^2 \to 0$ since $g_\rho\to 0$ at certain scale valid at some theoretically unknown high density $n_{vm}$, the hidden local gauge symmetry emerges in dense system. We will see below that $n_{vm}\gsim 25 n_0$ is indicated for the emergence of the pseudo-conformal sound velocity in stars.

Moreover, it was argued that the HLS fields could be (Seiberg-)dual to the gluons~\cite{Komargodski,kanetal,karasik}---the intrinsic quantity in QCD. At this moment, we do not know how could this happen. But, if this is right, we believe it means that the HLS gets {\it un-hidden} at high density. Since this duality indicates a Higgs phase-to-topological phase transition coinciding with the quark deconfinement at asymptotic density, it is most likely irrelevant to the compact stars we are concerned with~\cite{D} .

\subsection{Emergent scale symmetry}

It is well known that the scalar meson $f_0(500)$ is essential for providing the attractive force between nucleons. In our approach, it figures in as the Nambu-Goldstone boson of scale symmetry, the dilaton $\chi$. Actually, in Ref.~\cite{Schechter:1980ak}, the trace anomaly has been applied as a source of the scalar meson to construct the effective model of scalar meson by using anomaly matching. Here, we adopt the ``genuine dilaton" (GD) structure proposed in~\cite{crewther}.

The key premise of the GD idea is the existence of an infrared fixed point (IRFP) with the beta function $\beta (\alpha_{IR})=0$ for flavor number $N_f=3$ and we are living slightly away from this IRFP. Both the distance from the IRFP and current quark masses account for the dilaton mass. Explicitly, the dilaton mass is expressed as
\be
m_\sigma^2 f_\chi^2 = \langle \theta_\mu^\mu \rangle = \frac{\Delta_{IR} \beta^{\prime}(\alpha_{IR})}{\alpha_{IR}}\langle G_{\mu\nu}^a G^{a\mu\nu}\rangle + \cdots
\ee
where $\cdots$ stands for the contribution from quark mass and higher order of $\Delta_{IR} = \alpha_{IR} - \alpha_s$ with $\alpha_s$. This is in analogy to the Gell-Mann--Oakes-Renner relation in the pseudoscalar meson sector.

Since, unlike the unflavored hadrons, the effective masses of strange hadrons do not drop so much in dense medium, we will not consider the strangeness here.

Whether the proposed IRFP exists in QCD is still under debate. In Ref.~\cite{Alexandru:2019gdm},  it was argued  that in the IR region there is a nonperturbative  scale invariance different from that in the UV region. This is argued to lead to the possibility of massless glueballs in the fluid. What may be significant is the possible zero-mass glueball excitation. If we simply assume this picture works in dense system, this can be regarded as an indirect support of our theme. Anyway, we did not find any contradiction with  nature in using this GD idea.

\section{Pseudo-conformal model of compact star matter}

By using the G$n$EFT discussed above we are now in the position to calculate the nuclear matter properties.
We shall focus on the EOS of the baryonic matter, leaving out such basic issues as corrections to gravity, dark matters etc. Unless otherwise stated the role of leptons---electrons, muons, neutrinos, etc.,---is included in the EOS. Hereafter, we mainly focus on the effect of topology change. For other aspects, we refer to \cite{MR-review}.

\subsection{Density scaling}

In the construction of the PCM, we incorporate the medium modified hadron properties (dubbed as ``intrinsic density dependence (IDD)" to the G$n$EFT we constructed above by using the Brown-Rho scaling~\cite{Brown:1991kk} for $n\leq n_{1/2}$ (R-I) and the topological inputs for $n > n_{1/2}$ (R-II).

\sect{Density scaling in R-I}
In R-I, only one parameter $\Phi$ in Eq.~(\ref{scaling}) fixes all the IDDs. To the leading order in the chiral-scale counting~\cite{LMR}, the density scaling in R-I can be written as~\cite{MR-review}
\be
\frac{m^\ast_N}{m_N} \approx \frac{m^\ast_\chi}{m_\chi} \approx \frac{m^\ast_V}{m_V} \approx \frac{f^\ast_\pi}{f_\pi} \approx \frac{\langle \chi \rangle^\ast}{\langle \chi \rangle}\equiv \Phi(n), \label{scaling}
\ee
where $V=(\rho, \omega)$. Since there is no first-principle information on this quantity, for convenience, we fix it by taking the form
\be
\Phi_I=\frac{1}{1+c_I\frac{n}{n_0}}
\ee
with $c_I$ a constant. The range of $c_I$ that gives a good fit to nuclear matter properties as shown in Table.~\ref{table:parameter} is found to be~\cite{PKLR,PKLMR}
\be
c_I\approx 0.13-0.20\label{range}
\ee
with the upper value giving the measured pion decay constant~\cite{yamazaki-kienle}. Of course, it is expected as would be agreed by all nuclear physicists that certain fine-tuning  in the parameters be required for ground-state properties of nuclear matter.

\begin{table}[!htp]
\centering
\caption{Nuclear matter properties obtained at $n_0 < n_{1/2}$. The empirical values are merely exemplary. $n_0$ is in unit fm$^{-3}$ and others are in unit MeV.}
\label{table:parameter}
\label{Table}
\begin{tabular}{c|cc}
\hline
\hline
Parameter & Prediction  &Empirical  \cr
\hline
$n_0$ &~ $0.161$ &~ {$ 0.16\pm 0.01$~\cite{Pu:2017kjx}} ~ \cr
B.E. & ~ $16.7$ & ~ {$ 16.0\pm 1.0$~\cite{Pu:2017kjx}} ~ \cr
$E_{sym}(n_0)$ & ~ $30.2$ &~ $31.7\pm3.2$~\cite{Oertel:2016bki}~\cr
$E_{sym}(2n_0)$ & ~ $56.4$ &~ $46.9 \pm 10.1$~\cite{Li:2019xxz};$40.2 \pm 12.8 $~\cite{Chen:2015gba} ~\cr
{$L(n_0)$} & ~ {$67.8$} &~ {$58.9 \pm 16$~\cite{Li:2019xxz};$58.7\pm 28.1$~\cite{Oertel:2016bki}} ~\cr
$K_0$ & ~ $250.0$ &~ $ 230\pm20 $~\cite{Piekarewicz:2009gb} ~\cr
\hline
\hline
\end{tabular}
\end{table}

\sect{Density scaling in R-II}
Due to the topology change at $n_{1/2} > n_0$, the density dependence of some parameters are drastically different from that in R-I.

Since the hidden local gauge coupling $g_\rho$ and the $\rho$ meson mass are related to each other through the KSRF relation, we take the simplest form
\be
\frac{m_\rho^\ast}{m_\rho}\approx \frac{g_\rho^\ast}{g_\rho}\equiv \Phi_\rho\to \left(1- \frac{n}{n_{\rm VM}} \right)\ \ {\rm for } \ \ n > n_{1/2},\label{VMform}
\ee
where $n_{\rm VM}$ is the putative VM fixed-point density. How to join the $\Phi_\rho$ from $\Phi_I$ for $n\leq n_{1/2}$ is discussed in Ref.~\cite{PKLMR}. To have a result consistent with that from skyrmion crystal approach discussed above and mean field approach based on the leading order scale symmetry (LOSS)~\cite{Paeng:2013xya}, we take $n_{\rm VM} \gsim 25 n_0$.

The density scaling of the $\omega$ meson is more involved and different from that of $\rho$ meson which flows to the VM fixed point~\cite{Paeng:2011hy,PKLR}. It should be fine-tuned to match to the well constrained nuclear matter properties around the saturation density. Here, we take
\be
\frac{m_\omega^\ast}{m_\omega}\approx \kappa\frac{g_\omega^\ast}{g_\omega}
\ee
where $g_\omega$ is the $U(1)$ gauge coupling and
\be
\Phi_\omega\equiv \frac{g_\omega^\ast}{g_\omega}\approx 1-d\frac{n-n_{1/2}}{n_0}
\ee
with $d\approx 0.05$. This reflects the predicted break-down in R-II of the flavor $U(2)$ symmetry for the vector mesons which holds well  in R-I.

As for other parameters, we simply adopt the inputs from the skyrmion crystal approach, that is
\be
\frac{m_N^\ast}{m_N}\approx \frac{m_\sigma^\ast}{m_\sigma}\approx \frac{f_\chi^\ast}{f_\chi} \approx \frac{f_\pi^\ast}{f_\pi} \equiv \kappa\sim (0.6-0.9).\label{kappa}
\ee
The dilaton mass also goes proportional to the dilaton condensate. This follows from the partially conserved dilatation current (PCDC)~\cite{crewther}
\be
\frac{m_\sigma^\ast}{m_\sigma}\approx\kappa.
\ee
It follows also from low-energy theorems that
\be
\frac{g_{\pi NN}^\ast}{g_{\pi NN}}\approx \frac{g_A^\ast}{g_A}\approx\kappa.
\ee
The dilaton coupling to nucleon and other fields is unscaling to the leading order in scale-chiral symmetry, so it is a constant in R-II as in R-I.

\sect{$V_{lowk}$ renormalization group approach}
Equipped with the IDD, we are ready to calculate the EoS of nuclear matter. Here, to take into account the hadron fluctuation effects, we apply the $V_{lowk}$ renormalization group technique~\cite{Vlowk} which accounts for higher-order corrections to the Landau Fermi-liquid approximations~\cite{MR-review}. In this procedure, in addition to the IDD implemented in the density scaling of the parameters, the induced density dependence from the nucleon correlation denoted as DD$_{\rm induced}$\footnote{A typical example is the three-nucleon interaction in standard $\chi$EFT integrated into the scaling parameters.}  is also included. Therefore, the density dependence in the obtained EoS includes both IDD and DD$_{\rm induced}$. We denote the sum of IDD and DD$_{\rm induced}$ as $\overline{\rm IDD}$.

We would like to point out that, owing to the $\overline{\rm IDD}$ of the two-nucleon potentials, our calculation amounts to doing roughly N$^3$LO S$\chi$EFT including chiral 3-body potentials which are essential for the nuclear matter stabilized at the proper equilibrium density~~\cite{dongetal}. The same mechanism has been found to work for the C-14 dating Gamow-Teller matrix element where the three-body potential effect in S$\chi$EFT is reproduced by the $\overline{\rm IDD}$.

\subsection{The pseudo-conformal model of dense nuclear matter}

Using the density scaling discussed above, we can calculate the nuclear matter properties now. First we see from Table~\ref{table:parameter} that the empirical  values of the normal nuclear matter properties can be well reproduced.

Now, go to a higher density. Due to the topology change at $n_{1/2}$, there is a drastic change in the scaling of the parameters of G$n$EFT  leading to a qualitative impact on the structure of the EoS. So far, there is no theoretical argument to pin down $n_{1/2}$. Phenomenologically, we can estimate its range as $2.0n_0 < n_{1/2}<4.0n_0$ by using various astrophysical observations available, such as the maximum mass, the gravity-wave data and specially the star's  sound speed, and so on.

\begin{figure}[h]
\includegraphics[width=8cm]{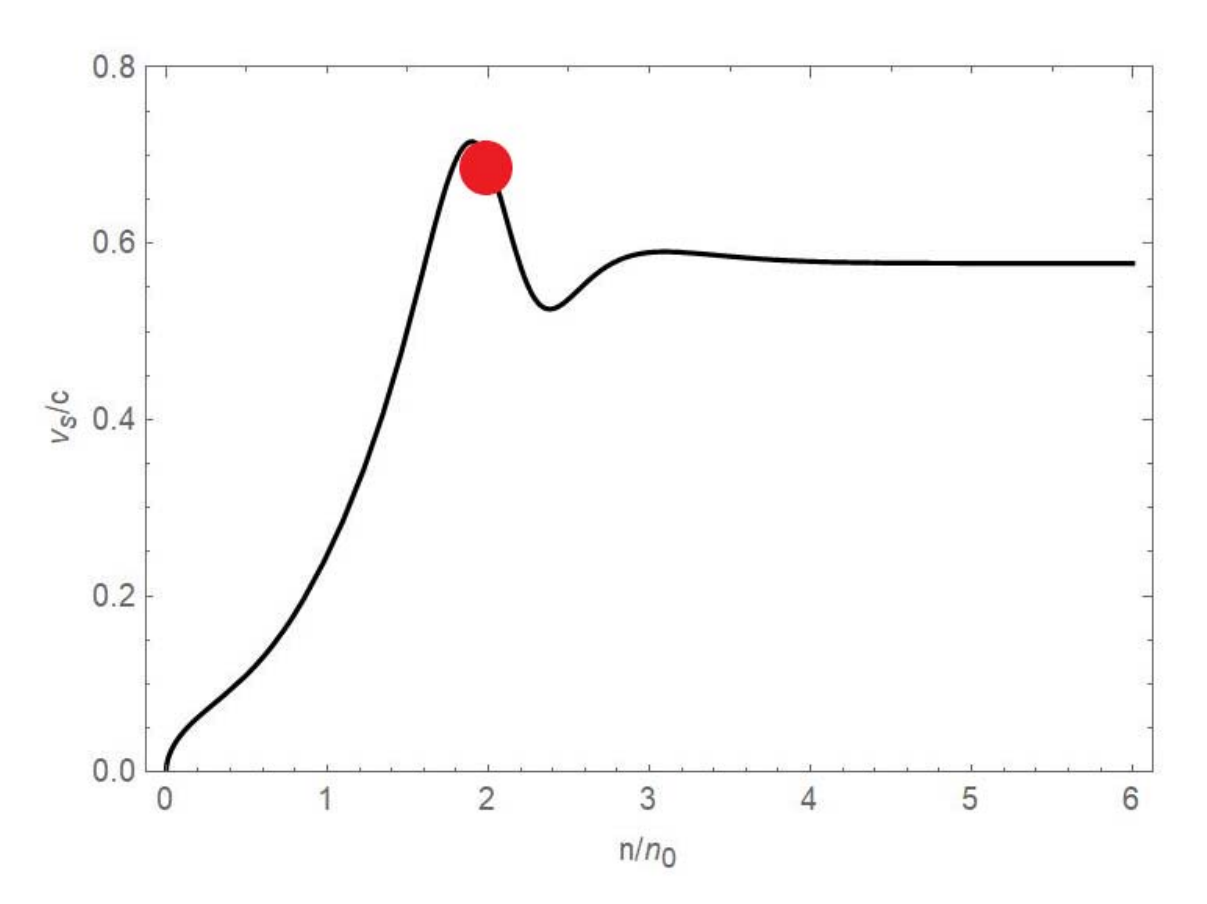}
\caption{Sound velocity calculated using $V_{lowk}$ RG with $n_{1/2}=2n_0$.  The solid circle indicates where the topology change takes place is.
}
\label{Vs}
\end{figure}

\sect{Sound velocity}
One of the most striking predictions that is in stark contrast to the conventional picture is the precocious appearance of the conformal sound velocity of the compact star matter. From Figure~\ref{Vs}, one can see that while the sound speed increases steadily and overshoots the conformal velocity at presumed $\sim n_{1/2}=2n_0$, it comes down and converges to $v_s^2\approx 1/3$.

It should be noted that the appearance of the conformal sound velocity at some high density is not so peculiar. Some reasonable $s\chi$EFT results resemble more or less this picture. But they show much broader and bigger bumps not exceeding the causality bound $v_s=1$ before converging to the conformal speed  $v_s^2=1/3$ but at an asymptotic density $\gsim 50n_0$~\cite{Tews:2018kmu}. After all, the convergence to the conformal speed at asymptotically high density is expected in perturbative QCD. What is striking and in a way unorthodox is the precocious onset of, and the convergence to, $v_s^2 \approx 1/3$ before reaching to an asymptotically high density despite that the trace of the energy-momentum tensor is nonzero. See below. It is somewhat like the ``quenched $g_A$" going to 1 in light nuclei~\cite{gA}, reflecting the pervasive imprint of hidden scale symmetry.

In our approach, the conformal sound speed follows as a logical outcome of the propositions~\cite{MR-review}, different from the parameter scanning done in \cite{Kanakis-Pegios:2020jnf}. These propositions yield that, going toward the DLFP~\cite{Paeng:2011hy}, the trace of  the energy-momentum tensor $\la\theta_\mu^\mu\ra$ is a function of only the dilaton condensate $\langle\chi\rangle^\ast$. Now if the condensate goes to a constant $\sim m_0$ due to the emergence of parity-doubling as we learned after the topology change, the $\la\theta_\mu^\mu\ra$ will become (more or less) independent of density. In this case, we will have
\be
\frac{\partial}{\partial n} \la\theta_\mu^\mu\ra=0.
\ee
This would imply that
\be
\frac{\partial\epsilon(n)}{\partial n}\left(1-3v_s^2\right)=0
\ee
where $v_s^2=\frac{\partial P(n)}{\partial n}/\frac{\partial\epsilon}{\partial n}$ and $\epsilon$ and $P$ are, respectively, the energy density and the pressure.
If we assume $\frac{\partial\epsilon(n)}{\partial n}\neq 0$, i.e., no Lee-Wick-type states in the range of densities involved, we can  then  conclude
\be
v_s^2=\frac{1}{3}.
\ee
This means that the dilaton condensate $\la\chi\ra^\ast $ goes to the density-independent constant $m_0$ due to the parity for $n_{\rm vm}\gsim 25 n_0$. This suggests the parity doubling at high density is linked to the $\rho$ decoupling from the nucleon together with the vector manifestation~\cite{Paeng:2011hy}.

\begin{figure}[h]
\begin{center}
\includegraphics[width=8.0cm]{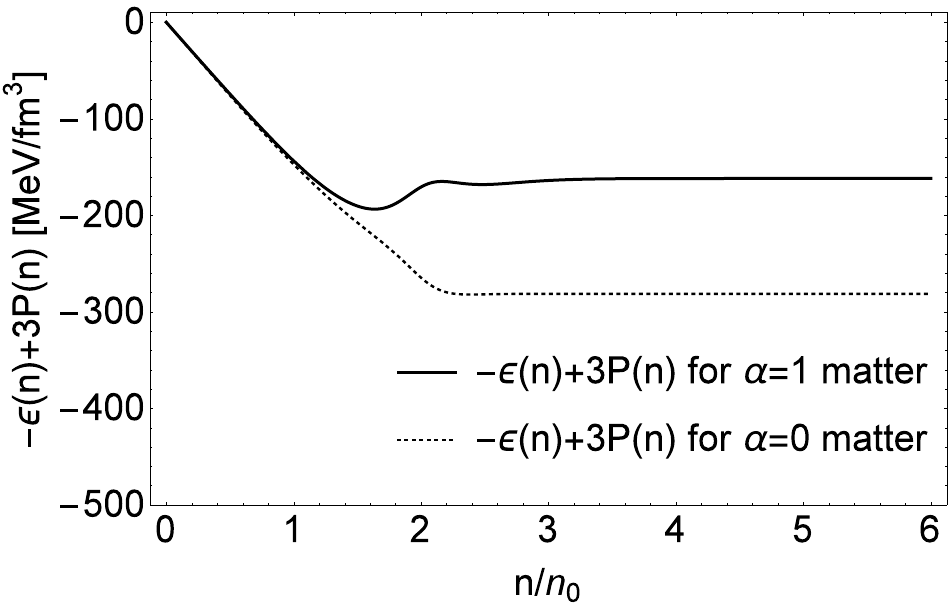}\;\;
\includegraphics[width=8.0cm]{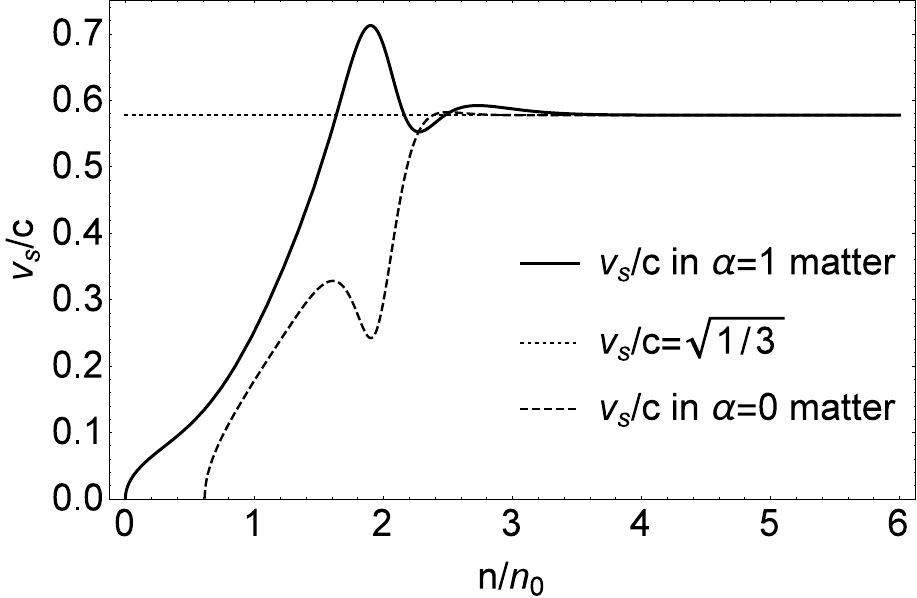}
\caption{$\la\theta_\mu^\mu\ra$ (upper panel) and $v_s$ (lower panel) vs. density for $\alpha=0$ (nuclear matter) and $\alpha=1$ (neutron matter) in $V_{lowk}$ RG for $n_{1/2}=2 n_0$.
 }\label{TEMT}
 \end{center}
\end{figure}

The above chain of reasoning is confirmed in the full $V_{lowk}$ RG formalism specifically for the case of $n_{1/2} = 2 n_0$. In Fig. \ref{TEMT} is shown the trace of the energy momentum tensor (left panel) that gives the conformal velocity for $n\gsim 3n_0$ (right panel).
This feature of both the TEMT and the sound velocity are expected to hold for any $n_{1/2}$ at which the topology change sets in,  i.e., within the range $2\lsim n_{1/2}/n_0\lsim  4 $.

\sect{Equation of state}
We now focus on the EoS of compact stars. It turns out that at density $n\geq n_{1/2}$, the conformality of the sound velocity can be captured by a simple two-parameter formula for the energy per-particle
\be
E/A= - m_N +X^\alpha  x^{1/3} + Y^\alpha x^{-1}\ \ {\rm with}\ x\equiv n/n_0
\label{PC-RII}
\ee
where $X$, $Y$ are parameters to be fixed. 

What we refer to as the pseudo-conformal model (PCM for short)  for the EoS is   then $E/A$ given by the union of that given by $V_{lowk}$ in R-I ($ n<n_{1/2}$) and that given by Eq.~(\ref{PC-RII}) in R-II ( $n\geq n_{1/2}$)  with the parameters $X^\alpha$ and $Y^\alpha$ fixed by the continuity at $n=n_{1/2}$ of the chemical potential and pressure
\be
\mu_I=\mu_{II},\ P_I=P_{II}\ \ {\rm at} \ \ n=n_{1/2}.
\label{matchingE}
\ee
This formulation is found to work very well for both $\alpha=0$ and 1 in the entire range of densities appropriate for massive compact stars, say up to $n\sim (6 - 7)n_0$, for the case $n_{1/2}=2n_0$ where the full $V_{lowk}$RG calculation is available~\cite{PKLMR}. We apply this PCM formalism for the cases where $n_{1/2}> 2n_0$.

Since a neutron star with mass $1.4M_\odot$ for which the tidal deformability $\Lambda$ obtained for $n_{1/2}=2.0n_0$ is $\Lambda_{1.4}\simeq 790$~\cite{PKLMR,PCM} that corresponds to the upper bound set by the gravity-wave data, we take the lower bound for the topology change density
\be
n_{1/2}\gsim 2n_0.\label{lower}
\ee

\begin{figure}[!h]
 \begin{center}
   \includegraphics[width=8cm]{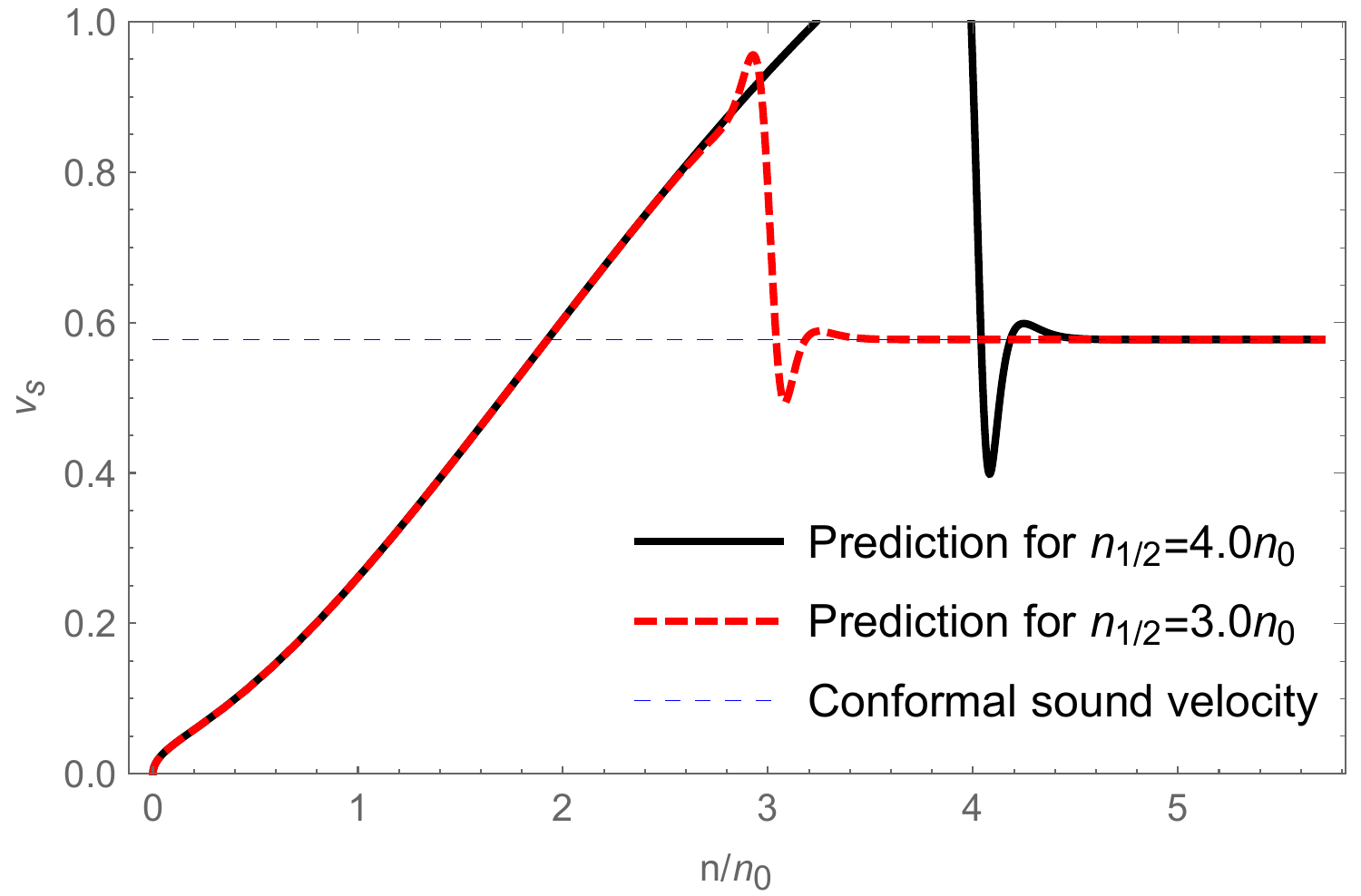}
  \end{center}
 \caption{Sound velocity as a function of density in neutron matter~\cite{Ma:2018qkg}.}
\label{Vs2-4}
\end{figure}

Next, let's see how the sound velocity comes out for $n_{1/2}/n_0=3\ {\rm and}\ 4$~\cite{PCM,Ma:2018qkg}. (The case for $n_{1/2}=2n_0$ was given in Fig.~\ref{TEMT}.) The results for neutron matter are summarized in Fig.~\ref{Vs2-4}.

It is clear from Fig.~\ref{Vs2-4} that, when $n_{1/2}=4 n_0$, the sound velocity violates the causality bound $v_s^2 < 1$.  The spike structure could very well be an artifact of the sharp connection made at the boundary. It may also be the different behaviour of the $\omega_0$ condensation at the low and high densities~\cite{Pisarski:2021aoz}. What is however physical is the rapid increase of the sound speed at the transition point signaling the changeover of the degrees of freedom.  Significantly, together with the lower bound (\ref{lower}), this allows us to pinpoint the region of the topology change
\be
2 n_0 \lsim  n_{1/2} \lsim  4 n_0.\label{boundhalf}
\ee
Later, we will explore whether or how  the waveforms of the gravitational waves emitted from the binary neutron star mergers respond  to  the location of $n_{1/2}$ which in our formulation corresponds to the point of hadron-quark continuity in QCD.

At this moment, we cannot obtain a more precise constraint than \eqref{boundhalf}. The important point is that it is an order of magnitude lower than the asymptotic density $\gsim 50 n_0$ that perturbative QCD predicts and signals the precocious emergence of pseudo-conformality in compact stars. However the robustness of the topological inputs figuring in the formulation convinces us that the precocious onset of the pseudo-conformal structure can be trusted at least qualitatively. In this connection, a recent detailed analysis of currently available data in the quarkyonic model is consistent with the possible onset density of $v_c^2\approx 1/3$ at $\sim 4 n_0$~\cite{lattimer}.

\begin{figure}[h]
 \begin{center}
   \includegraphics[width=8cm]{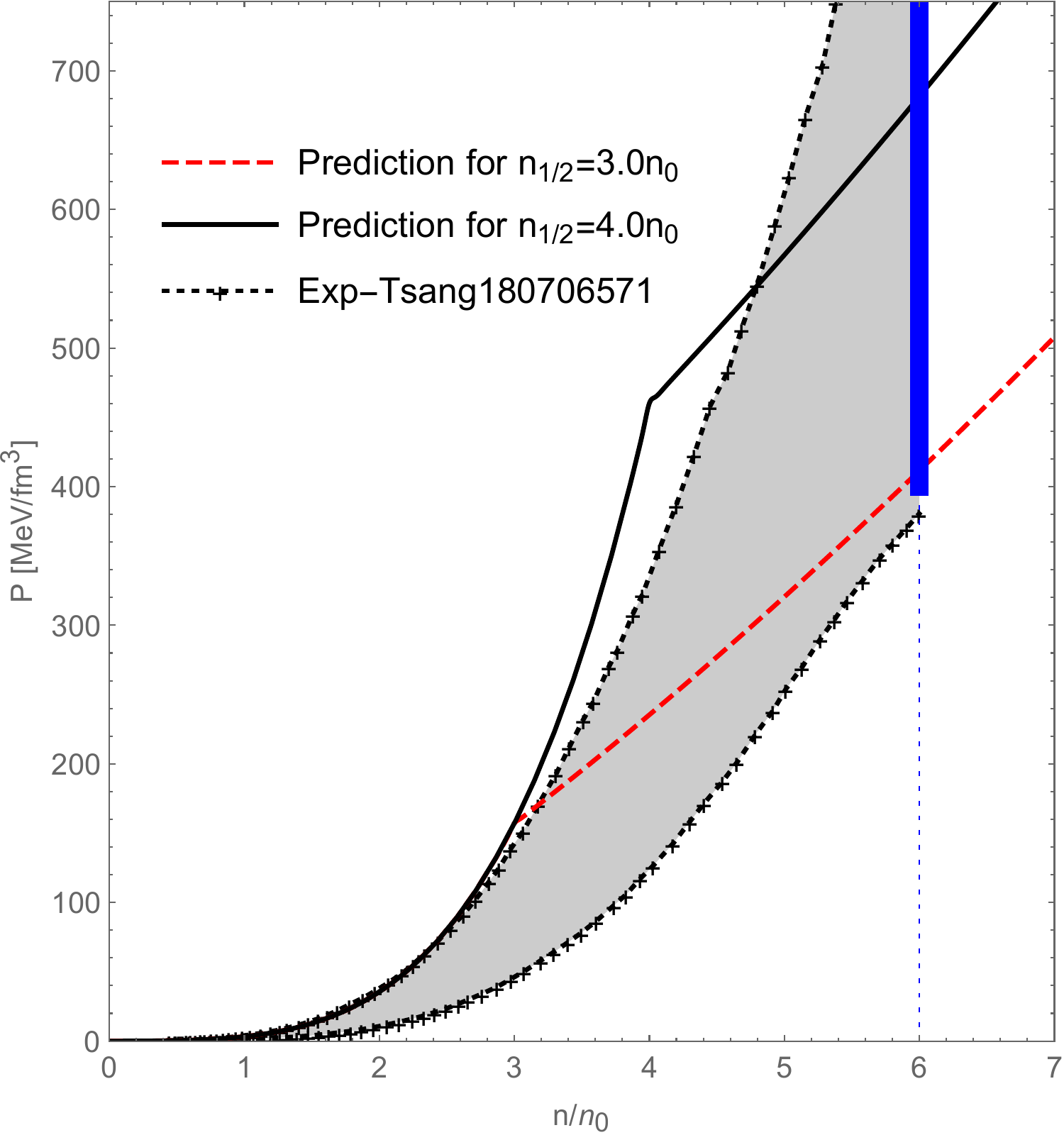}
  \end{center}
  \vskip -0.6cm
 \caption{Predicted pressure for neutron matter ($\alpha=1$) vs density compared  with the available experimental bound (shaded) given by Ref.~\cite{Tsang} and the bound at $6n_0$(blue band).}
\label{Tsang}
\end{figure}

Plotted in Fig.~\ref{Tsang} is the predicted pressure $P$ vs. density for $n_{1/2}/n_0= 3,4$ compared with the presently available heavy-ion data~\cite{Tsang}. The case of $n_{1/2}=4 n_0$, while consistent with the bound at $n\sim 6n_0$, goes outside of the  presently available experimental bound at $n\sim 4n_0$. This may again be an artifact of the sharp matching, but that it violates the causality bound seems to put it in tension with Nature. Nonetheless, without a better understanding of the cusp singularity present in the symmetry energy mentioned above it would be too hasty to rule out the threshold density $n_{1/2}=4n_0$.

\section{Star properties and gravitational waves}

\begin{figure}[h]
 \begin{center}
   \includegraphics[width=8cm]{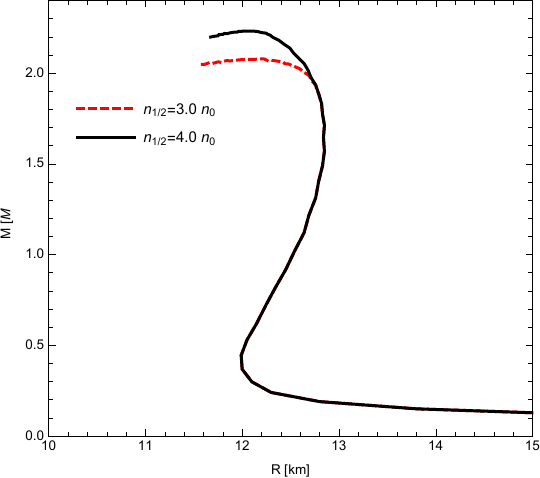} \includegraphics[width=8cm]{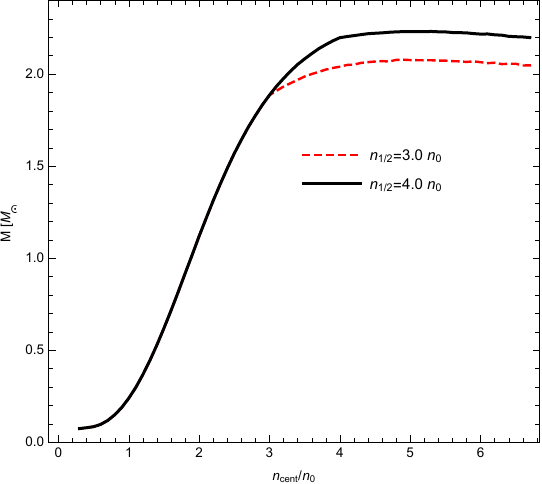}
 \end{center}
 \caption{Star mass $M$ vs. radius $R$ and central density $n_{\rm cent}$  with different choices of $n_{1/2}$. Note that below $M\approx 2M_\odot$, the curves for $n_{1/2}/n_0 =3.0\ {\rm and} \ 4.0$ represented in dashed line and solid line, respectively are coincident.}
\label{star-mass}
\end{figure}

\sect{Star mass}
The solution of the TOV equation with the pressures of leptons in beta equilibrium duly taken into account as in Ref.~\cite{PKLMR} yields the results for the star mass $M$ vs. the radius $R$ and the central density $n_{\rm cent}$ as given in Fig.~\ref{star-mass}. The maximum mass comes out to be roughly  $2.04M_\odot \sim 2.23 M_\odot$ for $2.0 \lesssim n_{1/2}/n_0 \lesssim 4.0$, the higher the $n_{1/2}$, the greater the maximum mass. This bound is consistent with the observation of the massive neutron stars
\be
M &=& 1.908\pm 0.016 M_\odot\ {\rm for}\ {\rm PSR}\ J1614-2230\mbox{~\cite{Demorest:2010bx}}
,\label{M3}\nonumber\\
&=& 2.01\pm 0.04 M_\odot\ {\rm for}\ {\rm PSR}\ J0348+0432\mbox{~\cite{Antoniadis:2013pzd}},\label{M2}\nonumber\\
&=& 2.14^{+0.10}_{-0.09} M_\odot\ \ {\rm for}\ {\rm PSR}\ J0740+6620\mbox{~\cite{Cromartie:2019kug}}.\label{M1}\nonumber
\ee
Note that this is not at odds with the conclusion of Ref.~\cite{Greif:2020pju} since in our model, the sound velocity exceeds the conformal limit in the intermediate density.

Fig.~\ref{star-mass} shows that, when $n_{1/2} \geq 3.0 n_0$, changing the position of $n_{1/2}$  affects only the compact stars with mass $\gtrsim 2.0 M_\odot$ although the mass-radius relation is affected by the topology change when $2.0 n_0 \leq n_{1/2} \leq 3.0 n_0$.

\sect{Tidal deformability}
Next, we confront our theory with what came out of the LIGO/Virgo gravitational observations---the dimensionless tidal deformability $\Lambda$. We will consider the dimensionless tidal deformability $\Lambda_i$ for the star $M_i$ and $\tilde{\Lambda}$ defined  by
\begin{eqnarray}
\tilde{\Lambda} & = &  \frac{16}{13}\frac{(M_1 + 12 M_2)M_1^4 \Lambda_1 + (M_2 + 12 M_1)M_2^4 \Lambda_2}{(M_1 + M_2)^{5}}
\label{eq:tildeL}
\end{eqnarray}
for $M_1$ and $M_2$ constrained to the well-measured ``chirp mass"
\begin{eqnarray}
{\cal M} = \frac{(M_1 M_2)^{3/5}}{(M_1 + M_2)^{1/5}} & = & 1.188 M_\odot ~~ \mbox{GW170817~\cite{Abbott:2018exr}},\nonumber\\
& = & 1.44 M_\odot ~~~ \mbox{GW190425~\cite{Abbott:2020uma}}.
\label{eq:chirpmass}
\ee

We plot our predictions for $\tilde{\Lambda}$ in Fig. \ref{lambdabar} and for $\Lambda_1$ vs. $\Lambda_2$ in Fig.~\ref{L1vsL2} {and compare our predictions with the results obtained with the parametrization of the EoS from the sound velocity constraints~\cite{Kanakis-Pegios:2020jnf}. As it stands, our prediction with $n_{1/2}\gsim 2n_0$ is compatible with the LIGO/Virgo constraint.  Although there seems to be some tension with the pressure, the result for $n_{1/2}=4 n_0$ is of quality comparable to that of $n_{1/2}=2 n_0$. A detailed analysis of the difference between PCM and ~\cite{Kanakis-Pegios:2020jnf} will be made later.}
\begin{figure}[h]
\vskip 0.5cm
 \begin{center}
\includegraphics[width=8cm]{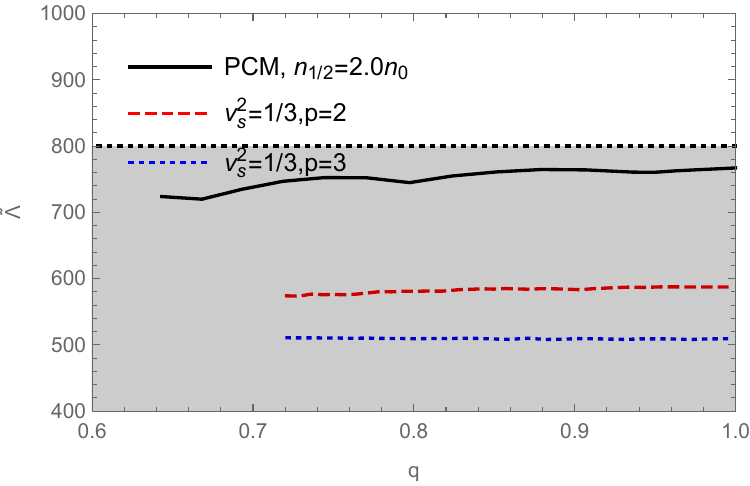}
\includegraphics[width=8cm]{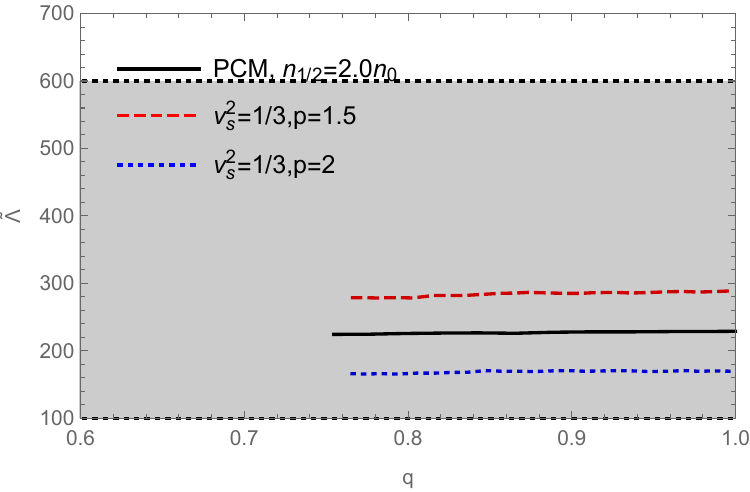}
  \end{center}
 \caption{The dimensionless tidal deformability $\tilde{\Lambda}$ as a function of the mass ratio $q$ for GW170817 with chirp
mass $1.188M_\odot$ (upper panel) and GW190425 with chirp mass $1.44M_\odot$ (lower panel). The the PCM prediction with $n_{1/2} = 2.0 n_0$ is plotted in solid line and those by  ~\cite{Kanakis-Pegios:2020jnf} are in dashed and dot-dashed lines (see Ref.~\cite{Kanakis-Pegios:2020jnf} for notation). The grey band in the upper panel is the constraint from the low spin
$\tilde{\Lambda} = 300^{+500}_{-190}$ obtained from GW170817~\cite{Abbott:2018exr} and that in the lower panel is $\tilde{\Lambda}\leq 600$ from GW190425.}
\label{lambdabar}
\end{figure}
\begin{figure}[h]
 \begin{center}
   \includegraphics[width=8cm]{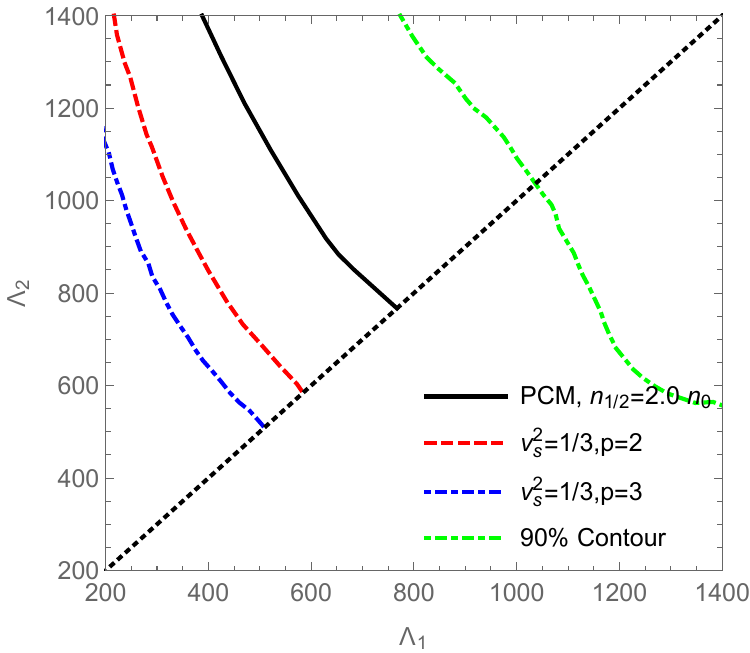}
   \includegraphics[width=8cm]{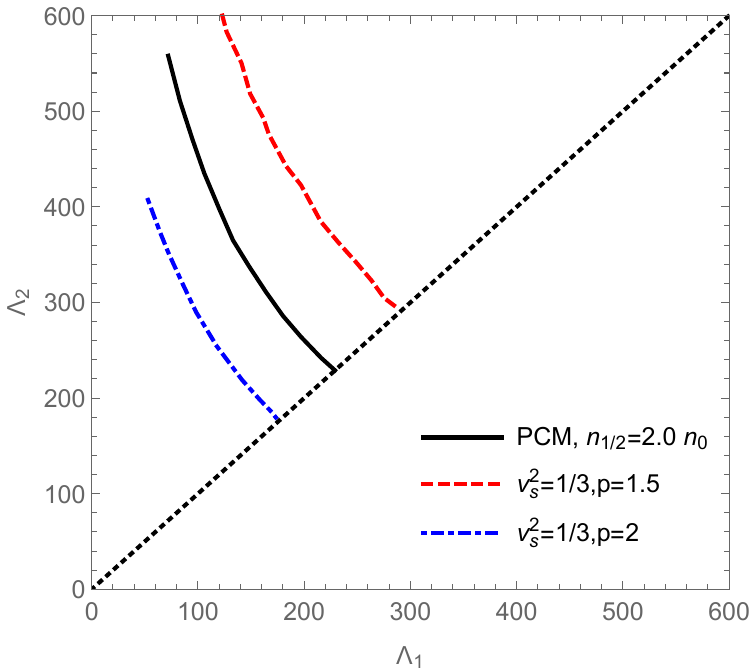}
  \end{center}
 %
\caption{Tidal deformabilities $\Lambda_1$ and $\Lambda_2$ associated with the high-mass $M_1$ and low mass $M_2$ components of the binary neutron star system GW170817 with chirp
mass $1.188M_\odot$ (upper panel) and GW190425 with chirp mass $1.44M_\odot$ (lower panel). The constraint from GW170817 at the 90\% probability contour is also indicated~\cite{Fattoyev:2017jql}.
}
\label{L1vsL2}
\end{figure}

\sect{Massive star composition}
Recently, combining astrophysical observations and model-independent theoretical {\it ab initio} calculations, Annala et al. arrive at the conclusion that the core of the massive stars is populated by  ``deconfined" quarks~\cite{evidence}. This is based on the observation that, in the core of the maximally massive stars, $v_s$ approaches the conformal limit $v_s/c \to 1/\sqrt{3}$ and the polytropic index takes the value $\gamma < 1.75$ --- the value close to the minimal one obtained in hadronic models.

We have seen above that, in the PCM, the predicted pseudo-conformal speed sets in precociously at $n \approx 3n_0$ and stays constant in the interior of the star. In addition, it is found that the polytropic index $\gamma$  drops, again rapidly, below 1.75 at $\sim 3n_0$ and approaches 1 at $n\gsim 6n_0$~\cite{Ma:2020hno}. This can be see from Fig.~\ref{vs}. Microscopic descriptions such as the quarkyonic model~\cite{quarkyonic} typically exhibit more complex structures at the putative hadron-quark transition density than our description, which is not unexpected given our picture is coarse-grained macroscopic description whereas the quarkyonic is a microscopic rendition of what's going on.

\begin{figure}[htbp]
\begin{center}
\includegraphics[width=0.45\textwidth]{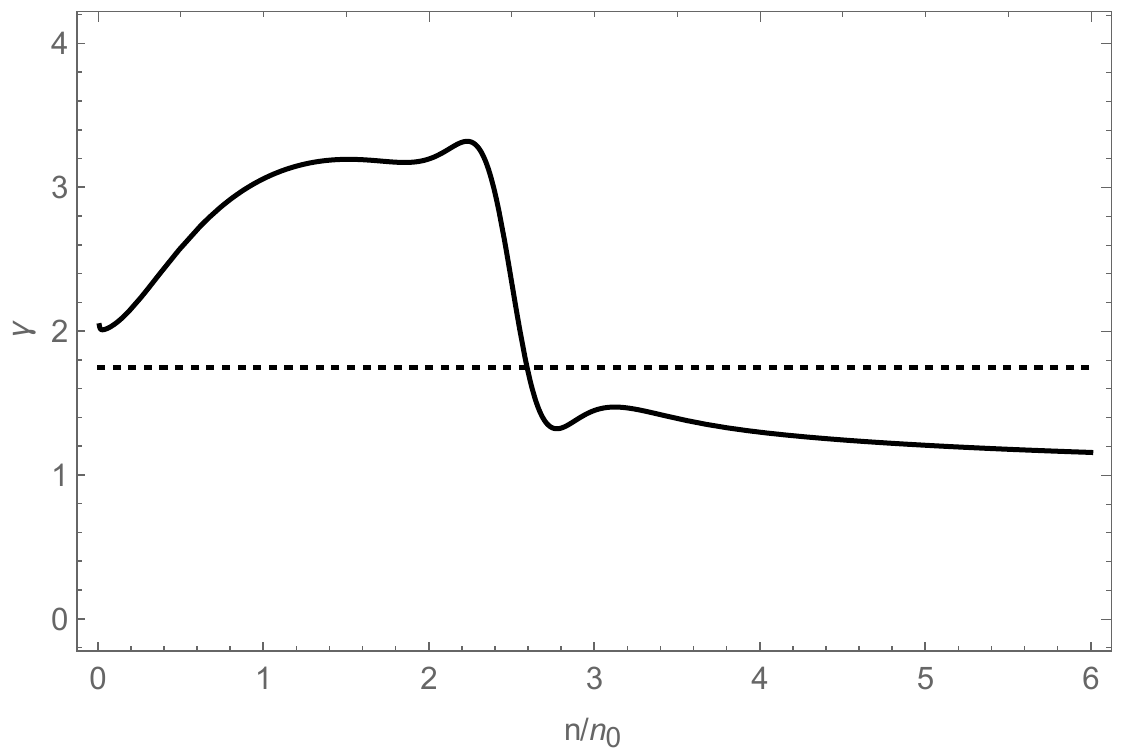}
\end{center}
\caption{Density dependence of the polytropic index $\gamma= {d\ln P}/{d\ln \epsilon}$ in neutron matter~\cite{Ma:2020hno}.}
\label{vs}
\end{figure}

To understand the origin of the similarity and difference between ~\cite{evidence} and PCM, we compare in Fig.~\ref{fig:EoS} our prediction for $P/\epsilon$ with the conformality band obtained by the SV interpolation method ~\cite{evidence}. We see that our prediction is close to, and parallel with, the conformality band. There are basic differences between the two. First of all, in our theory, conformality is broken---though perhaps only slightly at high density---in the system which can be seen from the deviation from the conformal band. Most importantly, the constituents of the matter after topology change in our theory is not (perturbatively) ``deconfined" quarks. It is a quasiparticle of fractional baryon charge, neither purely baryonic nor purely quarkonic. In fact it can be anyonic lying on a (2+1) dimensional sheet~\cite{D,trade-in}. That the predicted $P/\epsilon$ deviates from the conformal band is indicating that the scale symmetry the EoS of our theory is probing is some distance away from the IR fixed point with non-vanishing dilaton mass.

\begin{figure}[htbp]
\begin{center}
\vskip 0.3cm
\includegraphics[width=0.42\textwidth]{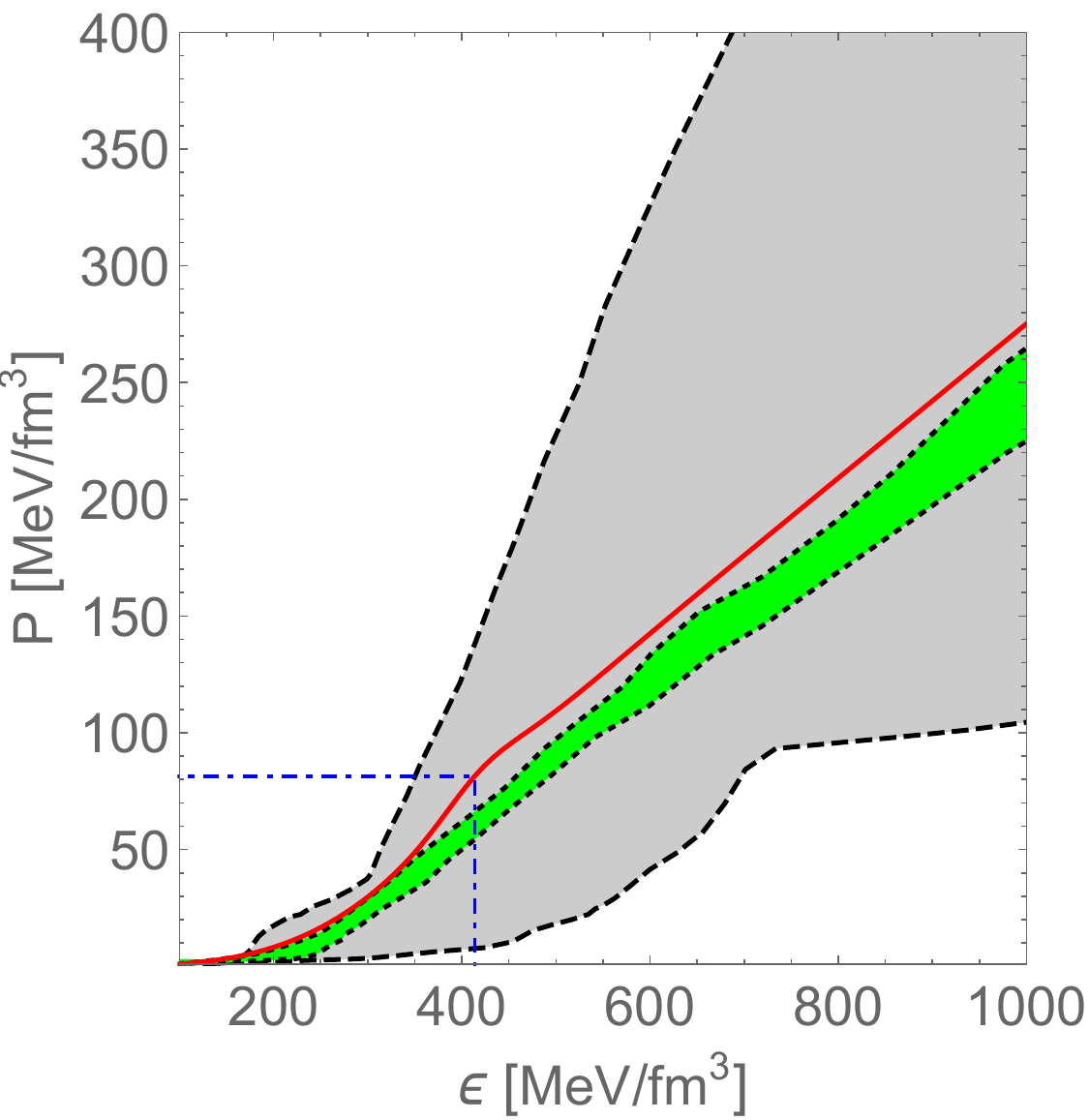}
\end{center}
\vskip -0.5cm
\caption{Comparison of $(P/\epsilon)$  between the PCM  velocity and the band from~\cite{evidence}. The gray band is  from the causality and the green band from the conformality. The red line is the PCM prediction. The dash-dotted line indicates the location of the topology change.}
\label{fig:EoS}
\end{figure}

\sect{Gravitational wave}
We finally apply our theory to the description of  the waveforms of the gravitational waves~\cite{Yang:2020awu}. The purpose is to explore whether one can probe the possible continuous crossover from hadrons to quarks represented in terms of the topology change. For this purpose, we consider the typical values $n_{1/2} = 2n_0$ and $3n_0$ and the neutron star mass $1.5M_\odot$.

The dominant mode of GW strain $h_{22}^+$ multiplied by the distance of the observer to the origin $R$ from BNS mergers is plotted in Fig.~\ref{fig:GW}. The plot shows  the location of the topology change affecting the number of the inspiral orbits, i.e., the number of the peaks in the inspiral phase which is the number of the peaks before merger, defined as the maximum of the amplitude of the GWs. Explicitly, the larger the $n_{1/2}$, the more the number of peaks. This could be within the detection ability of the on-going and up-coming facilities, especially the ground-based facilities~\cite{GWDetection}. The effect of the topology change on the waveforms can be understood from the distribution of the matter evolution of the BNS merger shown in Fig.~\ref{fig:matter}. It is found that the matter evolves faster when $n_{1/2} = 2n_0$ (the EoS is softer) than when  $n_{1/2} = 3n_0$ (the EoS is stiffer). Therefore the stars merge more easily with a shorter inspiral period. This indicates that  the waveforms of the gravitational waves emitted from the merger process as well as the matter evoluation could be sensitive to the EoS of compact stars (see, e.g., ~\cite{Hotokezaka:2013mm}). This observation explains the waveforms of Fig.~\ref{fig:GW}. However, there is a caveat: given that no qualitatively striking differences are predicted for all other astrophysical observables so far studied for $n_{1/2}/n_0=2$ and 3, it appears unnatural that the waveforms appear so different for only slightly different locations of the topology change. Furthermore since the transition involves no obvious phase change, at least  within the framework, the seemingly different impact of the topology change density---which is a coarse-grained description of the phenomenon---seems puzzling. It would be interesting to see whether the ``microscopic" models that simulate the quark degrees of freedom for hadron-quark continuity show similar sensitivity on the transition point. If indeed the waveforms were indeed very sensitive to the precise location of the cross-over, it would be extremely interesting.

\begin{figure}[tbp]
\includegraphics[width=3.2in]{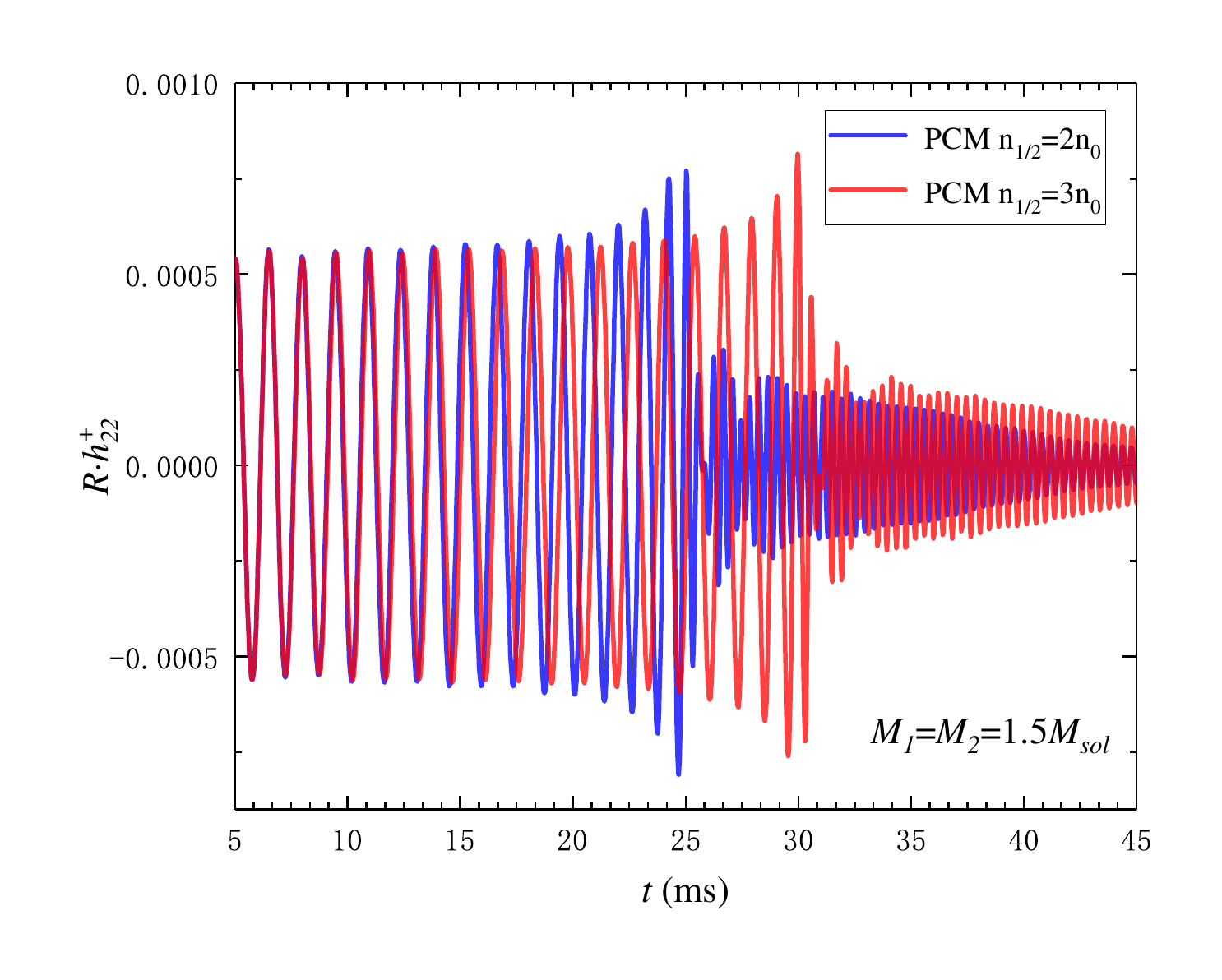}

	\caption{Dominant mode of the GW strain from the BNS star system with equal masses $1.5 M_{\odot}$ stars. For a detailed explanation, see Ref.~\cite{Yang:2020awu}.
}
	\label{fig:GW}
\end{figure}

\begin{figure*}[tbp]
	\centering
	\includegraphics[width=1.0\linewidth]{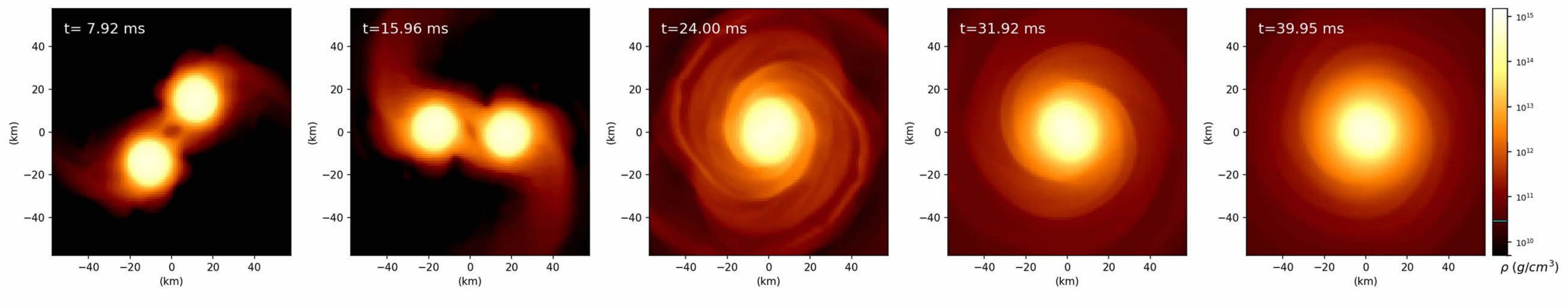}
    \includegraphics[width=1.0\linewidth]{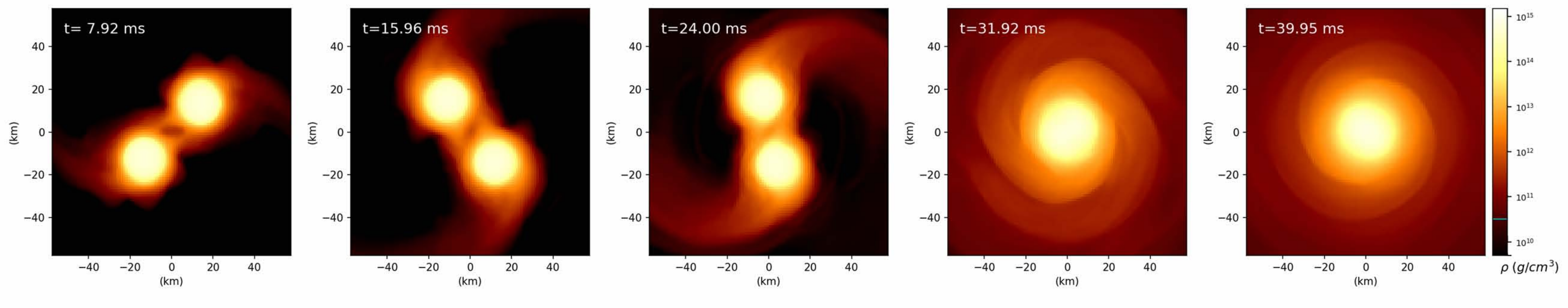}
	\caption{Matter density evolution of BNS mergers with equal mass $1.5M_\odot$ with (a) $n_{1/2}=2n_0$ (the upper row) and (b) $n_{1/2}=3n_{0}$ (the lower row)~\cite{Yang:2020awu}.
}
	\label{fig:matter}
\end{figure*}

\section{Summary and perspective}

In this work we reviewed the effect of the topology change representing the putative hadron-quark continuity on dense nuclear matter. The hadron and nuclear matter properties obtained from the skyrmion crystal approach, supplemented with the presumed emergent scale and flavor symmetries, inspired the construction of the pseudo-conformal model of dense nuclear matter relevant to compact stars. Locked to the density dependence of  hadron properties effected by the topology change at $n_{1/2}$, the trace of the energy momentum tensor of the model turns out to be a nonzero density-independent quantity and  induce the precocious appearance of the pseudo-conformal limit with $v_s^2 = 1/3$, in stark contrast to what's widely accepted in the field~\cite{Tews:2018kmu}.

So far, the pseudo-conformal model can describe the nuclear matter properties from low density to high density in a unified way. The nuclear matter properties calculated around the saturation density, the star properties such as the maximum mass, the mass-radius relation, the tidal deformability and so on all satisfy more or less satisfactorily the constraints from terrestrial experiments, astrophysical observations and gravitational wave detection.

Finally we state the possible caveats in and extensions of the model.

One can explicitly see from the above that although the tidal deformability predicted in the approach satisfies the currently cited constraint from the gravitational wave detection, it lies at the upper bound. But should the bound  turn out to go to a substantially lower value than what's given presently, the description of the cusp structure of the symmetry energy would need a serious revamping. In the present framework, the tidal deformability probes the density regime slightly below the  topology change density $n_{1/2}$ where the EoS is softer, and this is the density regime which is the hardest to control quantitatively in terms of the coarse-grained approach. It would require a more refined $V_{lowk}$-renormalization-group treatment than what has been done so far in \cite{PKLMR}, including the approximation made for the anomaly effect in the GD (genuine dilaton) scheme and the role of strangeness mentioned below.

One possible way to resolve the above caveat is to include the corrections to the LOSS, applied so far,  in such a way that, in addition to the mass parameters, the coupling constants also carry IDDs. This procedure may change the property of the EOS in the vicinity of $n_{1/2}$.  As a consequence the sound velocity after the topology change may also expose bumps, i.e., fluctuations from the conformal limit,  because of the explicit breaking of the conformal symmetry. However if the corrections from the explicit breaking of the conformal limit are taken as chiral-scale perturbation, the global picture of the compact star discussed would remain more or less intact.

Another point is the density at which the hidden scale and local flavor symmetries emerge. This is encoded in the IDDs of the hadron parameters such as pion decay constant, dilaton decay constant, $\rho$-N-N coupling and meson masses. By checking the effect of the location of the emergent symmetries on the star properties, one can also extract the information on the emergent symmetries and  the phase structure of QCD at low temperature.

Lastly we have left out the strangeness in the present discussion. It seems to have worked well without it in our approach up to now.  But there is of course no strong reason to ignore it. It could very well be that strangeness does play a crucial role but indirectly,  buried in the coarse-graining in the approach. Or it could also be that strangeness does not play a significant role up to the density involved in compact stars. The chiral-scale effective theory that  the pseudo-conformal description relies on is based on three flavor QCD with the scalar $f_0(500)$  taken on the same footing as the pseudoscalar mesons pion and kaon. There is however  a good reason to believe that in nuclear dynamics, the dilaton scalar is strongly affected by medium whereas the kaon is not. Implementing the strangeness in our approach would require doing the $V_{lowk}$ RG for 3-flavor systems with the hyperons treated on the Fermi sea   together with the nucleons as Fermi-liquid theory. This would then involve the kaon condensation as well as the hyperons as bound states of skyrmions and kaons. As argued in \cite{paeng-kaon}, it could postpone the role of strangeness to a much higher density than relevant to the most massive compact stars {\it stable against}  gravitational collapse. What happens beyond, such as color-flavor-locking, would be irrelevant to the problem.

One excuse for ignoring the strangeness could be that  the whole thing works without it, so why not adopt the spirit ``Damn the torpedoes! Full speed ahead!" \footnote{David Glasgow Farragut (1801-1870): ``Battle of Mobile Bay."} and proceed until hit by a torpedo?

%
{\bf Acknowledgments}

The work of Y.~L. M. was supported in part by National Science Foundation of China (NSFC) under Grant No. 11875147 and No.11475071.


{\bf Author's contributions}

All authors contributed equally to the review.

{\bf Funding}

Not Applicable.

{\bf Availability of data and materials}

No supplemental materials.

\section*{Declarations}
{\bf Ethics approval and consent to participate}

Not Applicable.

{\bf Consent for publication}

Not Applicable.

{\bf Competing interests}

There are no competing interests among the authors.


\end{document}